# Trends in risks of severe events and lengths of stay for COVID-19 hospitalisations in England over the pre-vaccination era: results from the Public Health England SARI-Watch surveillance scheme

*Peter D Kirwan[1], Suzanne Elgohari[2], Christopher H Jackson[1], Brian DM Tom[1], Sema Mandal[2], Daniela De Angelis[1,2], Anne M Presanis[1]*

[1] Medical Research Council Biostatistics Unit, School of Clinical Medicine, University of Cambridge, Cambridge, UK

[2] National Infection Service, Public Health England, London, UK

Corresponding author: peter.kirwan@mrc-bsu.cam.ac.uk

**Abstract**

**Background:** Trends in hospitalised case-fatality risk (HFR), risk of intensive care unit (ICU) admission and lengths of stay for patients hospitalised for COVID-19 in England over the pre-vaccination era are unknown.

**Methods:** Data on hospital and ICU admissions with COVID-19 at 31 NHS trusts in England were collected by Public Health England's Severe Acute Respiratory Infections surveillance system and linked to death information. We applied parametric multi-state mixture models, accounting for censored outcomes and regressing risks and times between events on month of admission, geography, and baseline characteristics.

**Findings:** 20,785 adults were admitted with COVID-19 in 2020. Between March and June/July/August estimated HFR reduced from 31.9% (95% confidence interval 30.3-33.5%) to 10.9% (9.4-12.7%), then rose steadily from 21.6% (18.4-25.5%) in September to 25.7% (23.0-29.2%) in December, with steeper increases among older patients, those with multi-morbidity and outside London/South of England. ICU admission risk reduced from 13.9% (12.8-15.2%) in March to 6.2% (5.3-7.1%) in May, rising to a high of 14.2% (11.1-17.2%) in September.

Median length of stay in non-critical care increased during 2020, from 6.6 to 12.3 days for those dying, and from 6.1 to 9.3 days for those discharged.

**Interpretation:** Initial improvements in patient outcomes, corresponding to developments in clinical practice, were not sustained throughout 2020, with HFR in December approaching the levels seen at the start of the pandemic, whilst median hospital stays have lengthened. The role of increased transmission, new variants, case-mix and hospital pressures in increasing COVID-19 severity requires urgent further investigation.

**Funding:** Medical Research Council.

## Introduction

COVID-19 patients admitted to hospital in England experience different possible pathways to a final outcome of either discharge or death, including admission to critical care (an intensive care unit (ICU) or high dependency unit (HDU)), a step-down from critical care to an acute ward, and transfers between hospitals. Knowledge of the risk of each pathway, the times taken (lengths of stay) to reach final outcomes and, particularly, how these change over time, are crucial to: (i) evaluate the effect of clinical and public health strategies on reducing disease burden (e.g. by reconstructing the evolution of the pandemic); and (ii) inform future policies, through medium- and long-term predictions of hospital demand [1,2]. Preliminary evidence exists for changes in mortality among critically ill patients with COVID-19 admitted to ICU/HDU in England during the first wave of the pandemic, with factors including age, sex, ethnicity and comorbidity associated with differing severity for hospitalised individuals [3–6]. However, little is known about how both risks of outcomes, and lengths of stay for all hospitalised patients have changed over time, particularly after the first pandemic wave up to the present vaccination era.

In March 2020, Public Health England (PHE) swiftly reconfigured existing influenza hospital surveillance to collect detailed individual-level data on COVID-19 patients, through the Severe Acute Respiratory Infections in England (SARI-Watch) surveillance system [7]. These data, which include all dates of movement between levels of care and outcomes and patient characteristics, allow estimation of the progression of COVID-19 patients through each hospital pathway. In this study we focus on how risks of each event and lengths of stay have varied over the months of the pandemic. In particular, we estimate changes in the hospital-fatality risk (HFR), averaged over the different pathways through hospital. Using parametric mixture multi-state models, we correctly account for both the competing risks of each pathway and missing outcome information (censoring) [8–10].

## Methods

### Study design and setting

Public Health England's (PHE's) Severe Acute Respiratory Infections in England (SARI-Watch) surveillance system, previously known as the COVID-19 Hospitalisations in England Surveillance System (CHESS), is a mandatory reporting system for all ICU/HDU admissions with COVID-19 at National Health Service (NHS) trusts in England [7]. A subset of 'sentinel' NHS trusts report information on the wider range of all hospital admissions.

### Participants

All adults aged 15+ admitted to 31 SARI-Watch sentinel NHS trusts between 15th March and 31st December 2020 with laboratory confirmed SARS-CoV-2 infection. Nosocomial cases of COVID-19 were excluded. Follow-up to death (COVID-19 associated or other cause) was extended until 11th January 2021, via linkage.

**Variables**

Variables collected in SARI-Watch include dates of hospital and critical care admission, date and type of outcome and baseline characteristics, specifically: age group, sex, region of residence (London, South, Midlands and East, and North of England), self-reported ethnicity and number of comorbidities. Due to small sample sizes, covariates were grouped where appropriate. Month groupings were: March, April, May, June/July/August, September, October, November and December; age groups were: 15-44, 45-64, 65-74, and 75+; region groupings were: London/South of England, Midlands and East of England, and North of England; ethnicity groupings were: Asian, Black, Mixed/Other and White; and comorbidity burden was grouped as: 0, 1, 2, and 3+ comorbidities. Comorbidity included pre-existing serious medical conditions such as: coronary heart disease, BMI≥30, diabetes, immunosuppression due to treatment or disease, renal disease, liver disease, neurological disorders, respiratory disease (excluding asthma), and asthma requiring medication.

**Data sources and outcomes**

SARI-Watch data are submitted to PHE via a secure online reporting tool, with data reporters able to update records with subsequent patient outcome information. Linkage to PHE's data on deaths among people with COVID-19 [11] was undertaken to obtain complete information on date and cause of COVID-19-related death. Self-declared ethnicity was obtained through linkage to the NHS Hospital Episodes Statistics database (HES) [12]. Linkage was undertaken for data up to 11th January 2021. Further details concerning these linkages are provided in the Appendix.

**Bias**

We obtained complete information on mortality and our statistical methodology was specifically designed to account for censored outcomes. The 31 sentinel trusts were verified to have consistent reporting of both critical and non-critical care patients over time and compared against aggregate admissions data for COVID-19 (see Appendix for list of trusts). Sentinel NHS trusts were originally recruited by PHE using regional and trust representation sampling, during the COVID-19 pandemic trusts chose whether to continue sentinel reporting leading to certain regions being over-represented, this may have introduced bias for which we could not account.

**Statistical methods**

We used a multi-state competing risks mixture model with four states: non-critical care, critical care (ICU), discharge and death, and two sub-models: events following hospital admission and events following critical care (ICU) admission. The design of this multi-state model is shown in Figure 1 and details of the model and model parameters are provided in the Appendix. Briefly, the model defines the probability that each competing event will occur next, and the distribution of the time to each event if that event occurs [13]. The probabilities of competing events were related to covariates through multinomial logistic regression, and the times to events were related to covariates through accelerated failure time models. The covariates considered included month of admission, age group, gender, ethnicity, region and number of comorbidities, as detailed in the Appendix. Time-to-event distributions and covariates on model parameters were chosen through critical assessment of goodness of fit (by comparison to non-parametric Aalen-Johansen cumulative incidence curves) and Akaike's information criterion (AIC).

The two sub-models shown in Figure 1 are a simplification of the patient pathway, to account for poorly completed information on unit type and critical care discharge dates. Critical care (ICU and HDU) admissions were combined into a single "ICU" state and additional hospital stays following ICU discharge were combined with the ICU state. Date of hospital admission, date of ICU admission for those admitted to ICU, and date and type of final outcome (discharge or death) were well completed.

**Censoring**

Overall, 13.7% of hospital admissions (2,843/20,785) and 17.7% of critical care admissions (392/2,213) had unknown final outcomes. These outcomes may be missing either because they were not reported despite having occurred (which we term "truly" missing), or because they had not occurred by the end of the period covered by the data (termed "right-censoring"). For earlier months, missing outcomes are more likely to be truly missing, whereas in the most recent months, a greater proportion of missing outcomes are likely to represent patients who remain in hospital care.

Among those with unknown outcome, a last known status was reported as either: 'still in hospital/still in ICU' at the date of data extraction, transferred on a particular date, or unknown. Patients 'still in hospital/still in ICU' were censored at the date of data extraction, or at 60 days if they were reported as still in non-critical care beyond 60 days, or at 90 days if they were reported as still in critical care beyond 90 days (Figure 2). This approach was a compromise between excluding these records from the analysis and censoring at the date of data extraction, which would allow for unrealistically long hospital stays of several months after admission. The 60/90-day windows included 99% of observed final outcomes, and 90-day in-hospital outcomes are also used in other studies of critical care [3]. Transferred patients with unknown outcome were censored at the date of transfer, patients with unknown status and unknown outcome were censored at the date of their last observed event.

**Model implementation**

Models were implemented using the flexsurv (version 2.0) package [14] in R (version 4.0.3) [15]. HFR, averaged over admissions with and without ICU, was estimated by combining the two sub-models. We report 95% confidence intervals representing uncertainty in estimates, and for lengths of stay we additionally report interquartile ranges representing heterogeneity within the population. Tables of estimated quantities, odds ratios and expected time ratios are provided in the Appendix.

**Results**

**Participant characteristics**

A total of 20,785 patients were admitted to a sentinel NHS trust between 15th March and 31st December 2020. The flowchart in Figure 2 shows the various pathways taken by patients and the aggregated number of observed transitions between each state.

Table 1 presents baseline characteristics by month of admission. Among those admitted to a sentinel NHS trust in March, 60% (2,037/3,391) were male, this proportion decreased to 50% (1,117/2,197) in May and remained between 50-55% for subsequent months. The age distribution of patients admitted to hospital fluctuated from month to month, with around 60% aged over 65 and 40% aged under 65. Meanwhile, the proportion of individuals of White ethnicity increased, from 63% (2,148/3,391) in March to 80% (1,768/2,200) in December, with a corresponding reduction in the proportion of patients of Asian, Black and Mixed/Other ethnicity. Patients admitted towards the end of the study period had fewer comorbidities; 44% (1,507/3,391) of patients admitted in March had three or more comorbidities, compared to 33% (721/2,200) of those admitted in December. The proportion with zero comorbidities rose from 18% (594/3,391) in March to 28% (623/2,200) in December. Differing regional trends were observed in hospital admissions; during March and April over half of admissions (60% and 54%, respectively) occurred in London and the South of England, whereas in September and October most hospital admissions (60% and 59%, respectively) occurred in the North of England.

**Critical care admissions**

Figure 3 shows the proportion of hospital admissions admitted to ICU over time by age and sex. Trends for non-critical care admissions were similar across all age groups and both sexes. A greater proportion of men compared to women and middle-aged (aged 45-75) patients compared to younger (aged 15-45) and older (aged 75+) patients were admitted to critical care.

**Outcome probabilities**

Estimated outcome probabilities following hospital admission are shown in Figure 4.

Probability of ICU admission following hospital admission dramatically decreased between March and May, from 13.9% (95% confidence interval 12.8-15.2%) to 6.2% (5.3-7.1%); subsequently rising to a peak of 14.2% (11.1-17.2%) in September, followed by a steady fall to 9.6% (8.6-10.9%) in December. These trends were similar across all subgroups.

Probability of death outside of critical care declined from 25.3% (23.8-26.8%) in March to 9.4% (7.9-11.1%) in June/July/August (OR 0.8, 0.6-1.1), followed by a resurgence to 23.1% (20.4-26%) in December (OR 1.3, 1.1-1.6). Over the study period, patients aged 65-74 and 75+ had significantly greater odds of death outside of critical care compared to patients aged 15-45 (OR 8.2, 3-8-18 and OR 71.1, 32.7-154.4, respectively). Patients with multiple comorbidities were similarly at greater risk of death outside of critical care (OR 2.3, 1.5-3.5 for 2 comorbidities and OR 5.2, 3.6-7.5 for 3+ comorbidities) compared to those with no comorbidities (Supplementary table 8).

Figure 5 shows estimated probability of death among patients admitted to critical care. This fell from 47.4% (42.5-51.7%) in March to 24.1% (16.1-35.1%) in June/July/August, increased to 40.2% (34.1-46.6%) in October and fell again to 26.8% (20.2-34.2%) in December. Whilst confidence intervals are wide in smaller groups, this oscillating trend was estimated across most subgroups.

**Hospitalised case-fatality risk**

Estimated hospitalised case-fatality risk (HFR), averaged across ICU and non-ICU admissions, is shown in Figure 6. Between March and June/July/August HFR reduced from 31.9% (30.3-33.5%) to 10.9% (9.4-12.7%); this was followed by an increase to 21.6% (18.4-25.5%) in September and steady rise to 25.7% (23.0-29.2%) for patients admitted in December. Whilst the initial fall in HFR was estimated across all subgroups, the increase to December was estimated particularly in the oldest age group (75+) and those with three or more comorbidities, with estimated HFRs of 42.6% (38.3-46.8%) and 34.5% (29.9-38.9%) for admissions in the final month of 2020, respectively.

In December, HFR was estimated to be higher in the North of England, at 28.6% (23.3-34.8%), and Midlands and East of England, at 27% (22.9-31.6%), compared to London and the South of England at 19.3% (16.1-23.3%). HFR for people of White ethnicity was higher during December, at 27.4% (24.3-30.8%), compared to 13% (9-19.3%) and 7.3% (3.3-18.6%) for people of Asian and Mixed/Other ethnicity, respectively. Whilst HFR for people of Black ethnicity was estimated to be lower than those of White ethnicity during March to June/July/August, confidence intervals were wider in December and overlapped the HFR for White ethnicity.

**Length of stay**

Estimated median times from hospital admission to next event are shown in Figure 7. Time to ICU admission remained low throughout the period, with a median of between 0.8 and 1.3 days after hospital admission (Figure 7A). Median length of stay in hospital prior to death in non-critical care increased from 6.6 days (6.2-7.1 days) in March to 12.3 days (10.9-14.0 days) in December, with a similar trend across all subgroups. Median lengths of stay in non-critical care prior to discharge initially fell from 6.1 days (5.7-6.5 days) for admissions during March to 4.1 days (3.8-4.5 days) in June/July/August, before lengthening to 9.3 days (8.5-10.3 days) in December. Heterogeneity between lengths of stay prior to discharge or death increased over time, with wider inter-quartile ranges estimated in more recent months.

Estimated length of stay in non-critical care prior to discharge was longest among older patients, with a median of 17.8 days (16.2-19.7 days) for patients aged 75+ (expected time ratio, ETR compared to 15-45: 3.7, 3.5-4) admitted in December (Figure 7C). Length of stay in non-critical care prior to discharge was also longer in the North (ETR compared to London/South 1.7, 1.6-1.9) compared to elsewhere in England; 13.8 days (12.3-15.6 days) in the North of England in December compared to 7.9 days (7.2-8.9 days) in London and the South of England and 9.9 days (9.0-11.1 days) in the Midlands and East of England (Figure 7D).

Figure 8 shows lengths of stay in ICU; as compared to non-critical care patients (Figure 7) patients admitted to critical care spent longer in hospital. Median length of stay in ICU prior to discharge initially reduced from 24.4 days (21.1-28.5 days) in March to 13.8 days (10.2-18.5 days) in June/July/August (ETR 0.6, 0.4-0.8), then increased to 38 days (29-48.9 days) in December (ETR 1.6, 1.2-2.1). Median length of stay in ICU prior to death showed less variability over time, ranging between 8.6 and 12.5 days.

## Discussion

### Principal findings

Since the start of the COVID-19 pandemic in England, risks of severe events and lengths of stay in hospital have varied by month of admission and according to baseline characteristics. The hospitalised case-fatality risk fell during the first 'wave' of the pandemic (March to June/July/August), although a resurgence was estimated across all subgroups up to December 2020, particularly among older patients, those presenting with multiple comorbidities, those of White ethnicity (unadjusted for age), and regions outside London and the South of England. Meanwhile, lengths of stay in hospital for patients admitted with COVID-19 have tended to increase since the start of the pandemic.

Temporal variations in patient outcomes and lengths of stay are likely to be influenced by a range of factors, including changes in patient characteristics and in clinical practice. Despite accounting for changes in patient characteristics over time, these covariates did not fully explain the decline and subsequent increase in severity for hospitalised COVID-19 patients, with trends persisting.

**Comparison with other studies**

Our estimates for critical care patients support the findings of the Intensive Care National Audit & Research Centre which reported a reduction in critical care mortality between March and June 2020, followed by an increase in mortality during the autumn and subsequent slow decline [3].

Considering all hospitalised patients, the International Severe Acute Respiratory and emerging Infections Consortium (ISARIC) estimated a reduction in in-hospital mortality among 208 hospitals in England, Scotland, and Wales, from 30-35% in March and early April to 10-15% in late July and August [6]. Meanwhile estimates of in-hospital mortality from the HES dataset, adjusted for age, sex, comorbidity, deprivation and date, showed a decline between early March and late May from 52.2% to 16.8% [16].

Our study, whilst limited to a selected group of hospitals, provides further evidence of this decline in hospitalised mortality during the first wave of COVID-19 in England. By accounting for competing risks, we additionally estimated trends in risks of ICU admission and hospital discharge, as well as lengths of stay for both critical and non-critical care patients. A key strength of our work is the longer follow-up period, which allowed for previously unexplored trends throughout 2020, in particular in autumn and winter, to be estimated.

Compared to the ISARIC study, we estimated a lower risk of ICU admission during March and April (17% vs 10 to 14%) [6]. Hospitals in London were the first to experience high numbers of critical care admissions during the first wave, so the fact that London sites were under-represented in the list of reporters to our study likely contributed to the reduced overall severity of infection in our estimates during these months compared to this study.

Several studies have shown association between increased severity due to COVID-19 and baseline covariates [3,5,6,16]. In common with these studies, we estimated increased risk of mortality for older patients and those with multiple comorbidities. These studies have also linked increased deprivation, and Asian, Black or Mixed/Other ethnicity compared to White ethnicity to increased severity. Full postcode information was unavailable and we were therefore unable to account for deprivation in our study. We estimated lower HFR for people of Asian, Black or Mixed/Other ethnicities compared to White ethnicity, this may be due to a differing age profile for these populations. Supplementary Figure 3 presents HFR by three covariates: month of admission, age group and ethnicity, showing that after controlling age, HFR was similar for people of Asian, Black and White ethnicities, although with wider confidence intervals in groups of smaller size.

**Strengths and limitations of the study**

By linking with PHE's mortality data we were able to identify all deaths due to COVID-19 occurring in the population, improving the completeness of the death information and thereby reducing uncertainty in our mortality estimates.

Another strength of this work is the statistical software we applied to surveillance data, allowing for robust estimates of outcome probability and lengths of stay to be obtained. Whilst we did not explicitly model seasonality, instead using a categorical month covariate, cyclic regression models or other smoothed functions of time could be used in future work to investigate seasonal variation in COVID-19 mortality rates.

Vaccination for COVID-19 in England began in mid-December 2020, with around 1.2 million people receiving their first dose by the end of December [17]. Our estimates cover the entire pre-vaccination period and therefore provide a comprehensive assessment of hospital severity before the effects of vaccination could be felt.

A limitation of the sentinel reporting was smaller group sizes compared to other studies [4,16], this led to increased uncertainty in our estimates, particularly among BAME patient groups, and limited analyses to only two covariates for the most part (month of admission plus one demographic factor). This limited the extent to which interactions between covariates could be explored, although it is likely that a high degree of correlation exists between certain covariates.

**Conclusions and policy implications**

Improved patient management, along with containment of COVID-19, is likely to have contributed to the estimated reduction in mortality during the first wave of infection in England [18]. From March onwards, hospital wards were re-configured to prepare for a rapidly increasing number of severe cases [19], nevertheless, several London-based hospitals entered 'surge', with availability of staff and ventilator units limiting the number of critical care beds [20]. At this time, there may have been prioritisation of ICU treatments to those most likely to benefit from invasive treatments, which may be reflected in our estimated improvements in patient outcomes. Concurrent national lockdown restrictions were effective in limiting transmission of COVID-19 in the community, slowing the influx of new patients from April onwards [21]. Use of treatments, such as Dexamethasone, and therapies such as prone positioning also improved outcomes for patients in non-critical care wards [22,23].

The estimated increase in mortality after the summer months may reflect a combination of multiple factors: increased transmission after easing of national restrictions; regional outbreaks [24]; changes in case-mix with a greater proportion of older, clinically vulnerable individuals observed being admitted to hospital [25]; seasonality; the emergence of new variants of concern (still a matter of debate and research [26]); and overloading of hospitals in hotspots and nationwide towards the end of the year [27]. Changes in critical care bed occupancy were worryingly correlated with estimated changes in mortality (Supplementary Figure 4), for example, although this crude comparison may hide a multitude of interlinked factors.

The concerning trends in hospitalised case-fatality risk and the relationship between increased transmission, appearance of new variants, hospital pressure, and mortality merits further urgent investigation. Changes in outcomes of patients admitted to hospital should continue to be closely monitored. Our results provide an invaluable baseline for future evaluation of the effect of vaccination on mortality and have been shared with the UK Government's Scientific Pandemic Influenza Group on Modelling (SPI-M) and Scientific Advisory Group for Emergencies (SAGE) [28,29].

**Acknowledgements:** We gratefully acknowledge the clinicians, data reporters and patients at SARI-Watch sentinel NHS trusts (listed in Appendix 1), as well as all Public Health England colleagues involved in the COVID-19 response.

**Patient and public involvement**

This study was a retrospective cohort analysis. The research question, design and data collection were motivated by the response to an urgent public health emergency. The surveillance data were collected by Public Health England with permissions granted under Regulation 3 of The Health Service (Control of Patient Information) Regulations 2002, and without explicit patient permission under Section 251 of the NHS Act 2006. Although patients were not directly involved in the study design, the experiences of clinicians and public health officials interacting with patients informed the design of the data collection.

**Data sharing:** Requests to access non-publicly available data are handled by the PHE Office for Data Release (ODR). (https://www.gov.uk/government/publications/accessing-public-health-england-data/about-the-phe-odr-and-accessing-data)

**Dissemination to participants and related patient and public communities:** PHE and the MRC Biostatistics Unit have public facing websites and Twitter accounts @PHE_uk and @MRC_BSU. PHE and the MRC Biostatistics Unit engage with print and internet press, television, radio, news, and documentary programme makers.

**Funding:** This research is funded by the Medical Research Council (PK, DDA, CJ, AP, Unit programme number MC_UU_00002/11, BT Unit programme number MC_UU_00002/2); and via a grant from the MRC UKRI/ DHSC NIHR COVID-19 rapid response call (AP, DDA, BT, grant ref: MC_PC_19074). This research is also supported by the NIHR Cambridge Biomedical Research Centre.

**Role of the funding source:** The funders had no influence on the methods, interpretation of results or decision to submit.

**Ethics approval:** SARI-Watch is a mandatory surveillance system approved by the Department of Health and Social Care. Data were collected with permissions granted under Regulation 3 of The Health Service (Control of Patient Information) Regulations 2002, and without explicit patient permission under Section 251 of the NHS Act 2006.

**Transparency statement:** The lead author affirms that this manuscript is an honest, accurate, and transparent account of the study being reported; that no important aspects of the study have been omitted; and that any discrepancies from the study as planned (and, if relevant, registered) have been explained.

**Authors' contributions:** All authors conceived the research study. PK and AP drafted the manuscript. CJ developed the flexsurv package and advised on methods. PK carried out the multi-state model analyses. SE co-ordinated the linkage to ONS and HES and formatted the dataset. The dataset was verified by PK and CJ. All authors provided critical review of the manuscript prior to submission. The corresponding author attests that all listed authors meet authorship criteria and that no others meeting the criteria have been omitted. PK is the guarantor.

**Competing interests statement:** All authors have completed the Unified Competing Interest form (available on request from the corresponding author) and declare: no support for the submitted work from anyone other than their employer, no financial relationships with any organisations that might have an interest in the submitted work in the previous three years, no other relationships or activities that could appear to have influenced the submitted work.

**References**

1.  Birrell PJ, Blake J, van Leeuwen E, Cell PJM, Gent N, De Angelis D. Real-time Nowcasting and Forecasting of COVID-19 Dynamics in England: the first wave? medRxiv [Internet]. 2020; Available from: https://www.medrxiv.org/content/10.1101/2020.08.24.20180737v1.abstract

2.  Scientific Pandemic Influenza Group on Modelling, Operational sub-group. SPI-M-O Medium-Term Projections [Internet]. 2021 Feb 17. Available from: https://assets.publishing.service.gov.uk/government/uploads/system/uploads/attachment_data/file/966763/S1120_SPI-M-O_MediumTermProjections.pdf

3.  Intensive Care National Audit & Research Centre. ICNARC report on COVID-19 in critical care: England, Wales and Northern Ireland [Internet]. 2021 Feb. Available from: https://www.icnarc.org/Our-Audit/Audits/Cmp/Reports

4.  Ferrando-Vivas P, Doidge J, Thomas K, Gould DW, Mouncey P, Shankar-Hari M, Young JD, Rowan KM, Harrison DA, ICNARC COVID-19 Team. Prognostic Factors for 30-Day Mortality in Critically Ill Patients With Coronavirus Disease 2019: An Observational Cohort Study. Crit Care Med. 2021 Jan 1;49(1):102–11.

5.  Williamson EJ, Walker AJ, Bhaskaran K, Bacon S, Bates C, Morton CE, Curtis HJ, Mehrkar A, Evans D, Inglesby P, Cockburn J, McDonald HI, MacKenna B, Tomlinson L, Douglas IJ, Rentsch CT, Mathur R, Wong AYS, Grieve R, Harrison D, Forbes H, Schultze A, Croker R, Parry J, Hester F, Harper S, Perera R, Evans SJW, Smeeth L, Goldacre B. Factors associated with COVID-19-related death using OpenSAFELY. Nature. 2020 Aug;584(7821):430–6.

6.  Docherty AB, Mulholland RH, Lone NI, Cheyne CP, De Angelis D, Diaz-Ordaz K, Donoghue C, Drake TM, Dunning J, Funk S, García-Fiñana M, Girvan M, Hardwick HE, Harrison J, Ho A, Hughes DM, Keogh RH, Kirwan PD, Leeming G, Nguyen-Van-Tam JS, Pius R, Russell CD, Spencer R, Tom BDM, Turtle L, Openshaw PJM, Baillie JK, Harrison EM, Semple MG, for ISARIC4C investigators. Changes in UK hospital mortality in the first wave of COVID-19: the ISARIC WHO Clinical Characterisation Protocol prospective multicentre observational cohort study [Internet]. bioRxiv. medRxiv; 2020. Available from: http://medrxiv.org/lookup/doi/10.1101/2020.12.19.20248559

7.  Public Health England. Weekly national Influenza and COVID-19 surveillance report [Internet]. 2021 Feb. Report No.: week 6 report (up to week 5 data). Available from: https://www.gov.uk/government/statistics/national-flu-and-covid-19-surveillance-reports

8.  Andersen PK, Geskus RB, de Witte T, Putter H. Competing risks in epidemiology: possibilities and pitfalls. Int J Epidemiol. 2012 Jun;41(3):861–70.

9.  Jackson CH, Tom BDM, Kirwan PD, Seaman SM, Kunzmann K, Presanis AM, De Angelis D. A comparison of two frameworks for multi-state modelling, applied to outcomes after hospital admissions with COVID-19. Submitted. 2021;

10. Aalen O, Borgan O, Gjessing H. Survival and Event History Analysis: A Process Point of View. Springer; 2008. 540 p.

11. Covid-19 EpiCell. PHE data series on deaths in people with COVID-19: technical summary [Internet]. Public Health England; 2020 Aug. Available from: https://www.gov.uk/government/publications/phe-data-series-on-deaths-in-people-with-covid-19-technical-summary

12. NHS Digital. Hospital Episode Statistics (HES) [Internet]. 2020 [cited 2021 Feb 24]. Available from: https://digital.nhs.uk/data-and-information/data-tools-and-services/data-

services/hospital-episode-statistics

13. Larson MG, Dinse GE. A mixture model for the regression analysis of competing risks data. J R Stat Soc [Internet]. 1985; Available from: https://rss.onlinelibrary.wiley.com/doi/abs/10.2307/2347464

14. Jackson CH. flexsurv: A Platform for Parametric Survival Modeling in R. J Stat Softw. 2016 May 12;70(8):1–33.

15. R Core Team. R: A Language and Environment for Statistical Computing [Internet]. R Foundation for Statistical Computing, Vienna, Austria; 2020. Available from: https://www.R-project.org/

16. Navaratnam AV, Gray WK, Day J, Wendon J, Briggs TWR. Patient factors and temporal trends associated with COVID-19 in-hospital mortality in England: an observational study using administrative data. Lancet Respir Med [Internet]. 2021 Feb 15; Available from: http://dx.doi.org/10.1016/S2213-2600(20)30579-8

17. UK Government. Official UK Coronavirus Dashboard [Internet]. 2021 [cited 2021 Feb 24]. Available from: https://coronavirus.data.gov.uk/

18. Public Health England. COVID-19: investigation and initial clinical management of possible cases [Internet]. Guidance. 2020 [cited 2021 Feb 18]. Available from: https://www.gov.uk/government/publications/wuhan-novel-coronavirus-initial-investigation-of-possible-cases/investigation-and-initial-clinical-management-of-possible-cases-of-wuhan-novel-coronavirus-wn-cov-infection

19. Stevens S, Pritchard A. Urgent next steps on NHS response to COVID-19 [Internet]. 2020. Available from: https://www.england.nhs.uk/coronavirus/wp-content/uploads/sites/52/2020/03/urgent-next-steps-on-nhs-response-to-covid-19-letter-simon-stevens.pdf

20. National Health Service. Clinical guide for the management of surge during the Coronavirus pandemic: critical care rapid learning [Internet]. 2020 Nov. Available from: https://www.nice.org.uk/Media/Default/About/COVID-19/Specialty-guides/management-of-surge.pdf

21. Davies NG, Barnard RC, Jarvis CI, Russell TW, Semple MG, Jit M, Edmunds WJ, Centre for Mathematical Modelling of Infectious Diseases COVID-19 Working Group, ISARIC4C investigators. Association of tiered restrictions and a second lockdown with COVID-19 deaths and hospital admissions in England: a modelling study. Lancet Infect Dis [Internet]. 2020 Dec 23; Available from: http://dx.doi.org/10.1016/S1473-3099(20)30984-1

22. Recovery Collaborative Group. Dexamethasone in hospitalized patients with Covid-19—preliminary report. New England Journal of Medicine [Internet]. 2020; Available from: https://www.nejm.org/doi/full/10.1056/NEJMoa2021436

23. Bamford P, Bentley A, Dean J, Whitmore D, Wilson-Baig N. ICS guidance for prone positioning of the conscious COVID patient 2020. London: Intensive Care Society, 2020 [Internet]. 2020. Available from: https://emcrit.org/wp-content/uploads/2020/04/2020-04-12-Guidance-for-conscious-proning.pdf

24. Nazareth J, Minhas JS, Jenkins DR, Sahota A, Khunti K, Haldar P, Pareek M. Early lessons from a second COVID-19 lockdown in Leicester, UK. Lancet. 2020 Jul 18;396(10245):e4–5.

25. National Health Service. Clinical guide for the management of emergency department patients

during the coronavirus pandemic [Internet]. 2020 Nov. Available from: https://www.nice.org.uk/media/default/about/covid-19/specialty-guides/management-emergency-department.pdf

26. Scientific Advisory Group for Emergencies. NERVTAG: Update note on B.1.1.7 severity, 11 February 2021 [Internet]. GOV.UK; 2021 [cited 2021 Feb 24]. Available from: https://www.gov.uk/government/publications/nervtag-update-note-on-b117-severity-11-february-2021

27. Roxby BP, Butcher B, England R. Pressure on hospitals "at a really dangerous point." BBC [Internet]. 2020 Dec 18 [cited 2021 Feb 18]; Available from: https://www.bbc.co.uk/news/health-55362681

28. GOV.UK. Scientific pandemic influenza group on modelling (SPI-M) [Internet]. GOV.UK; 2020 [cited 2021 Feb 24]. Available from: https://www.gov.uk/government/groups/scientific-pandemic-influenza-subgroup-on-modelling

29. GOV.UK. Scientific Advisory Group for emergencies [Internet]. GOV.UK; 2020 [cited 2021 Feb 24]. Available from: https://www.gov.uk/government/groups/scientific-advisory-group-for-emergencies-sage-coronavirus-covid-19-response

**Tables**

Table 1: Baseline characteristics of patients with COVID-19 admitted to sentinel NHS trusts during 15th March to 31st December 2020, by month and covariate

| Group | Mar | Apr | May | Jun/Jul/Aug | Sep | Oct | Nov | Dec |
|---|---|---|---|---|---|---|---|---|
| All | 3391 (100%) | 6923 (100%) | 2197 (100%) | 1304 (100%) | 450 (100%) | 1861 (100%) | 2459 (100%) | 2200 (100%) |
| **Sex** | | | | | | | | |
| Female | 1354 (39.9%) | 2969 (42.9%) | 1080 (49.2%) | 619 (47.5%) | 203 (45.1%) | 851 (45.7%) | 1168 (47.5%) | 1045 (47.5%) |
| Male | 2037 (60.1%) | 3954 (57.1%) | 1117 (50.8%) | 685 (52.5%) | 247 (54.9%) | 1010 (54.3%) | 1291 (52.5%) | 1155 (52.5%) |
| **Age group** | | | | | | | | |
| 15-44 | 414 (12.2%) | 718 (10.4%) | 296 (13.5%) | 229 (17.6%) | 90 (20%) | 220 (11.8%) | 337 (13.7%) | 339 (15.4%) |
| 45-64 | 1023 (30.2%) | 1973 (28.5%) | 524 (23.9%) | 382 (29.3%) | 132 (29.3%) | 488 (26.2%) | 599 (24.4%) | 619 (28.1%) |
| 65-74 | 610 (18%) | 1237 (17.9%) | 363 (16.5%) | 240 (18.4%) | 75 (16.7%) | 403 (21.7%) | 482 (19.6%) | 361 (16.4%) |
| 75+ | 1344 (39.6%) | 2995 (43.3%) | 1014 (46.2%) | 453 (34.7%) | 153 (34%) | 750 (40.3%) | 1041 (42.3%) | 881 (40%) |
| **Region of residence** | | | | | | | | |
| London/South of England | 2020 (59.6%) | 3712 (53.6%) | 783 (35.6%) | 457 (35%) | 55 (12.2%) | 191 (10.3%) | 462 (18.8%) | 635 (28.9%) |
| Midlands and East of England | 911 (26.9%) | 1914 (27.6%) | 731 (33.3%) | 436 (33.4%) | 127 (28.2%) | 570 (30.6%) | 1039 (42.3%) | 1072 (48.7%) |
| North of England | 460 (13.6%) | 1297 (18.7%) | 683 (31.1%) | 411 (31.5%) | 268 (59.6%) | 1100 (59.1%) | 958 (39%) | 493 (22.4%) |
| **Ethnicity** | | | | | | | | |
| Asian | 424 (12.5%) | 848 (12.2%) | 184 (8.4%) | 158 (12.1%) | 72 (16%) | 152 (8.2%) | 189 (7.7%) | 184 (8.4%) |
| Black | 520 (15.3%) | 653 (9.4%) | 124 (5.6%) | 83 (6.4%) | 11 (2.4%) | 25 (1.3%) | 28 (1.1%) | 55 (2.5%) |
| Mixed/Other | 163 (4.8%) | 320 (4.6%) | 81 (3.7%) | 49 (3.8%) | 15 (3.3%) | 27 (1.5%) | 53 (2.2%) | 68 (3.1%) |
| White | 2148 (63.3%) | 4799 (69.3%) | 1730 (78.7%) | 975 (74.8%) | 338 (75.1%) | 1573 (84.5%) | 2106 (85.6%) | 1768 (80.4%) |
| Unreported* | 136 (4%) | 303 (4.4%) | 78 (3.6%) | 39 (3%) | 14 (3.1%) | 84 (4.5%) | 83 (3.4%) | 125 (5.7%) |
| **Number of comorbidities** | | | | | | | | |
| 0 | 594 (17.5%) | 1402 (20.3%) | 383 (17.4%) | 291 (22.3%) | 121 (26.9%) | 416 (22.4%) | 603 (24.5%) | 623 (28.3%) |
| 1 | 629 (18.5%) | 1246 (18%) | 406 (18.5%) | 270 (20.7%) | 114 (25.3%) | 347 (18.6%) | 481 (19.6%) | 459 (20.9%) |
| 2 | 661 (19.5%) | 1284 (18.5%) | 410 (18.7%) | 215 (16.5%) | 80 (17.8%) | 368 (19.8%) | 483 (19.6%) | 397 (18%) |
| 3+ | 1507 (44.4%) | 2991 (43.2%) | 998 (45.4%) | 528 (40.5%) | 135 (30%) | 730 (39.2%) | 892 (36.3%) | 721 (32.8%) |

* Unreported ethnicity excluded from multi-state mixture model.

**Figures**

Figure 1: Multi-state model for COVID-19 hospital admissions. Arrows indicate transitions between states in each of the 'From Hospital' and 'From ICU' sub-models.

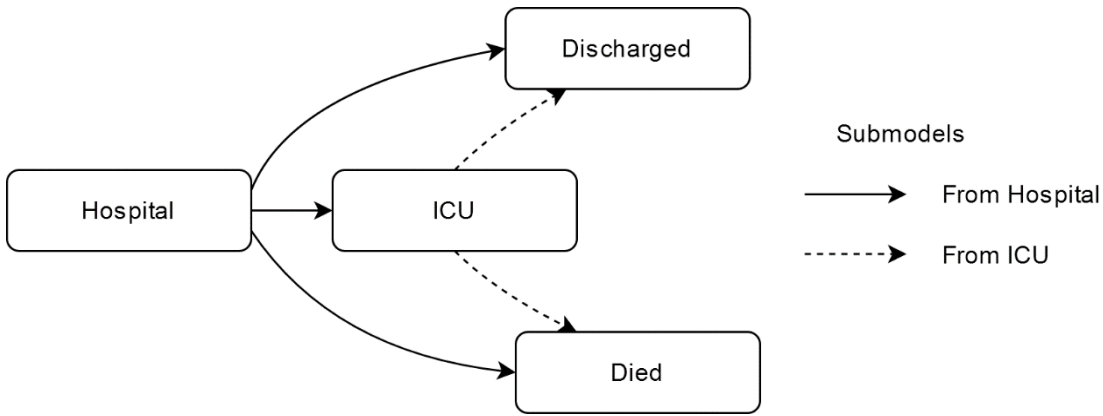

Figure 2: Flow diagram showing pathways through multi-state model, with observed number and proportion of individuals in and moving between each state. Dotted lines indicate right-censoring, i.e. those with unknown or not yet experienced outcomes. Dashed lines indicate right-censoring for individuals reported as "still in hospital" for >60 days or "still in ICU" for >90 days.

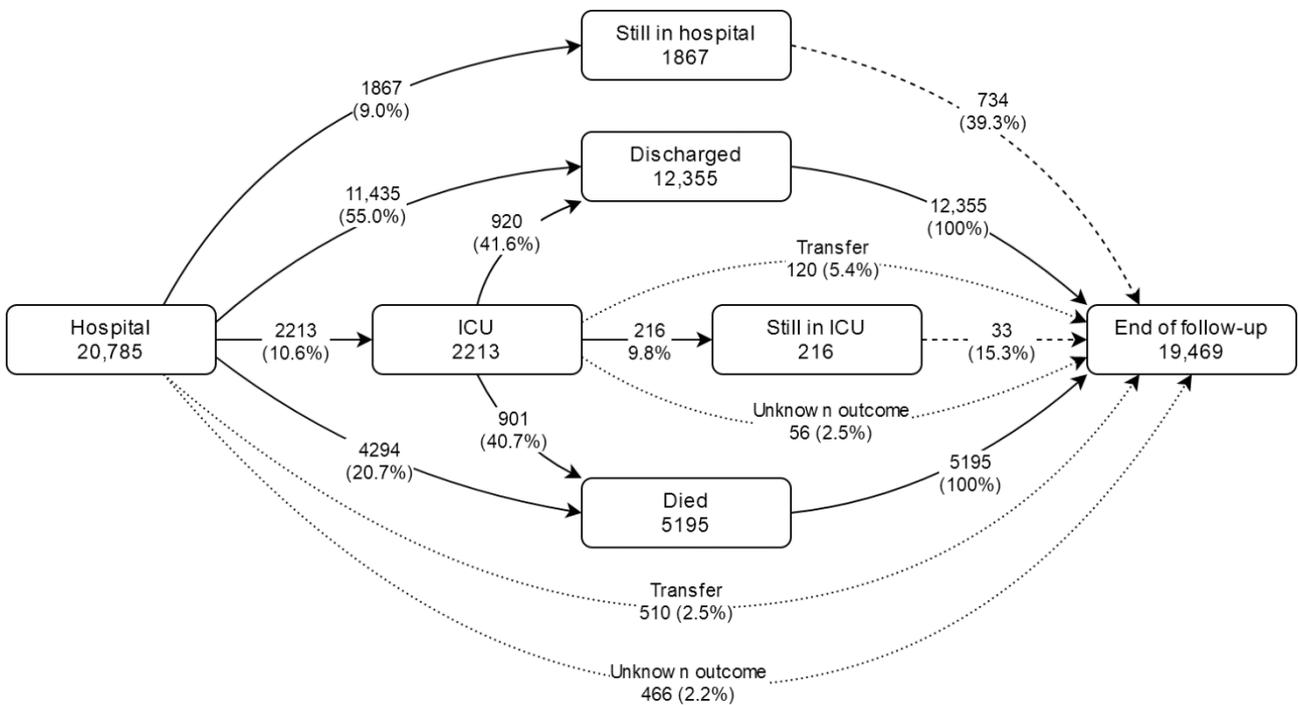

Figure 3: Proportion of COVID-19 hospital admissions by ICU admission status, gender and age group.

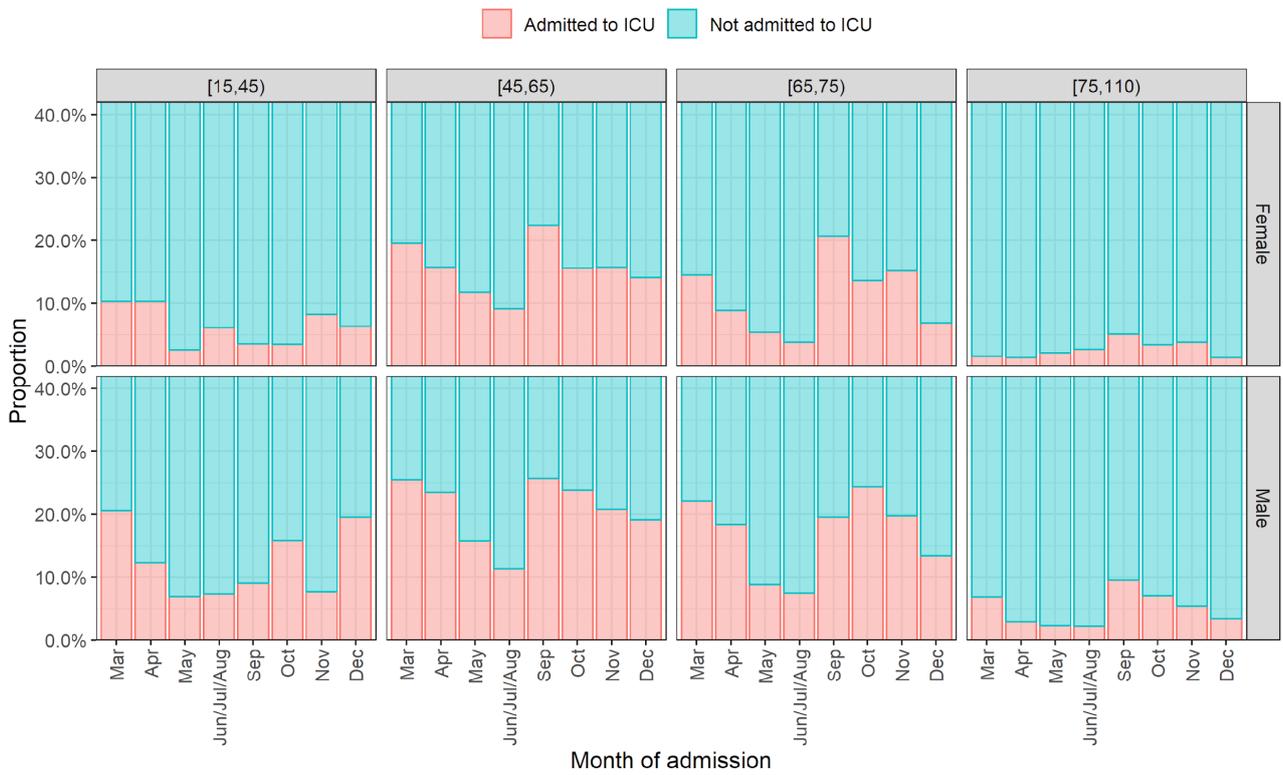

Figure 4: Probability of next event following hospital admission ('From Hospital' sub-model), by month of admission (panel A), sex (panel B), age group (panel C), region of residence (panel D), ethnicity (panel E), and number of comorbidities (panel F). Error bars are 95% confidence intervals to represent uncertainty in the estimated probability.

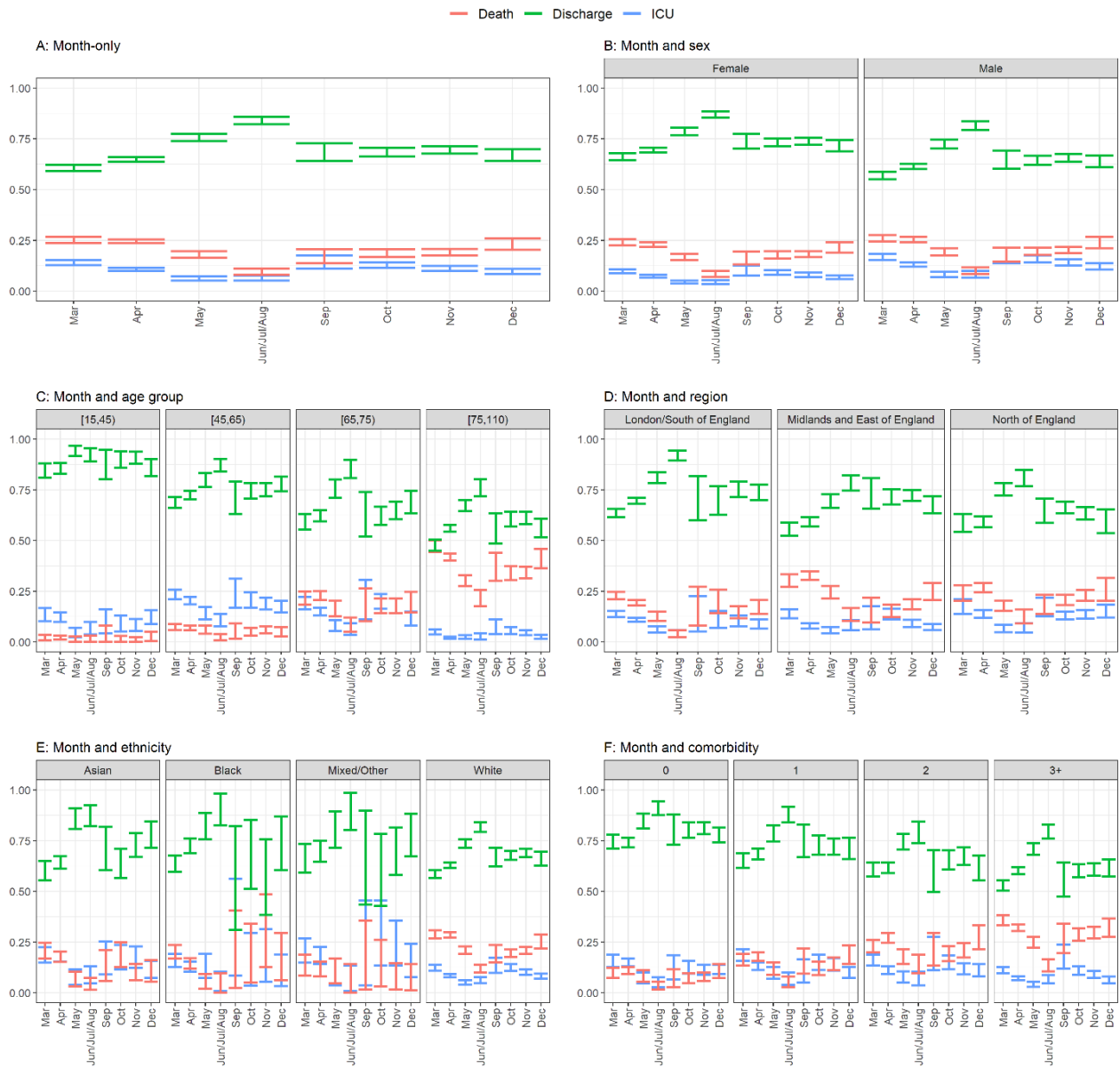

Figure 5: Probability of death following ICU admission ('From ICU' sub-model), by month of admission (panel A), sex (panel B), age group (panel C), region of residence (panel D), ethnicity (panel E), and number of comorbidities (panel F). Error bars are 95% confidence intervals to represent uncertainty in the estimated probability.

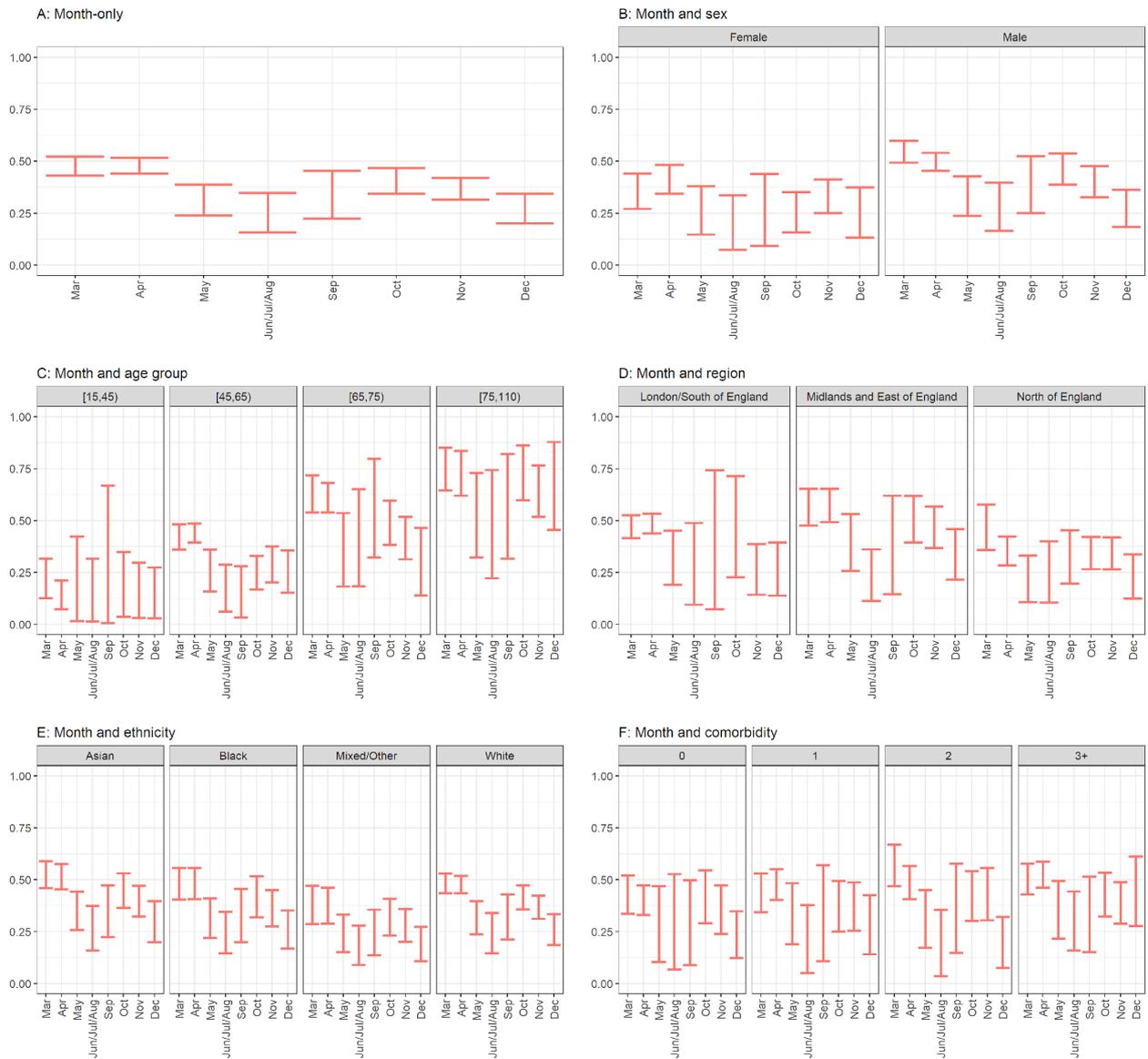

Figure 6: Estimated hospitalised case-fatality risk (HFR) averaged over ICU and non-ICU admission, by month of admission (panel A), sex (panel B), age group (panel C), region of residence (panel D), ethnicity (panel E), and number of comorbidities (panel F). Y-axis constrained to 0-0.6. Error bars are 95% confidence intervals to represent uncertainty in the estimated probability.

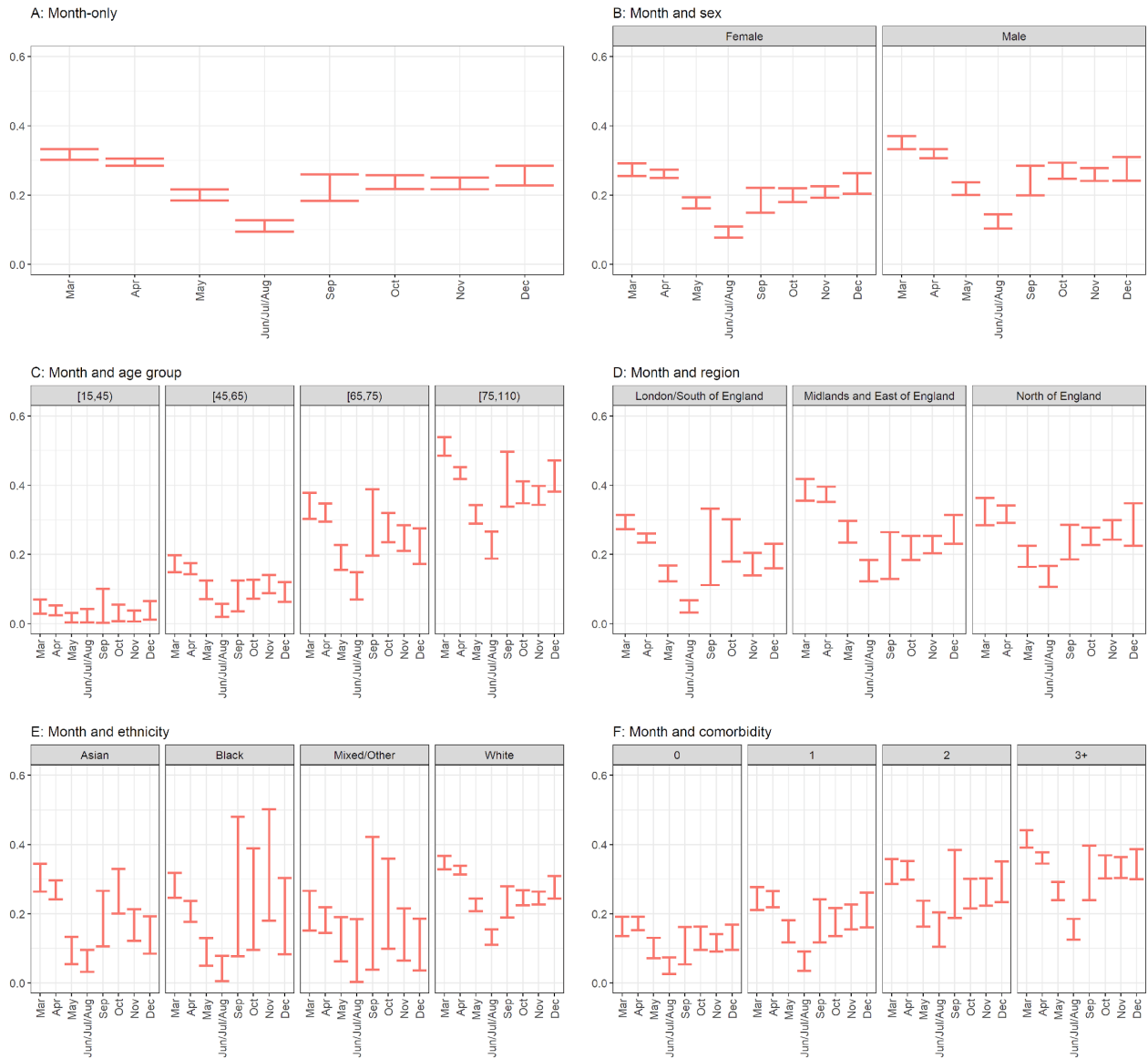

Figure 7: Estimated median time to event following hospital admission ('From Hospital' sub-model), by month of admission (panel A), sex (panel B), age group (panel C), region of residence (panel D), ethnicity (panel E), and number of comorbidities (panel F). Line ranges are interquartile ranges, representing variability in estimated times to event, error bars are 95% confidence intervals, representing uncertainty in the estimated median time.

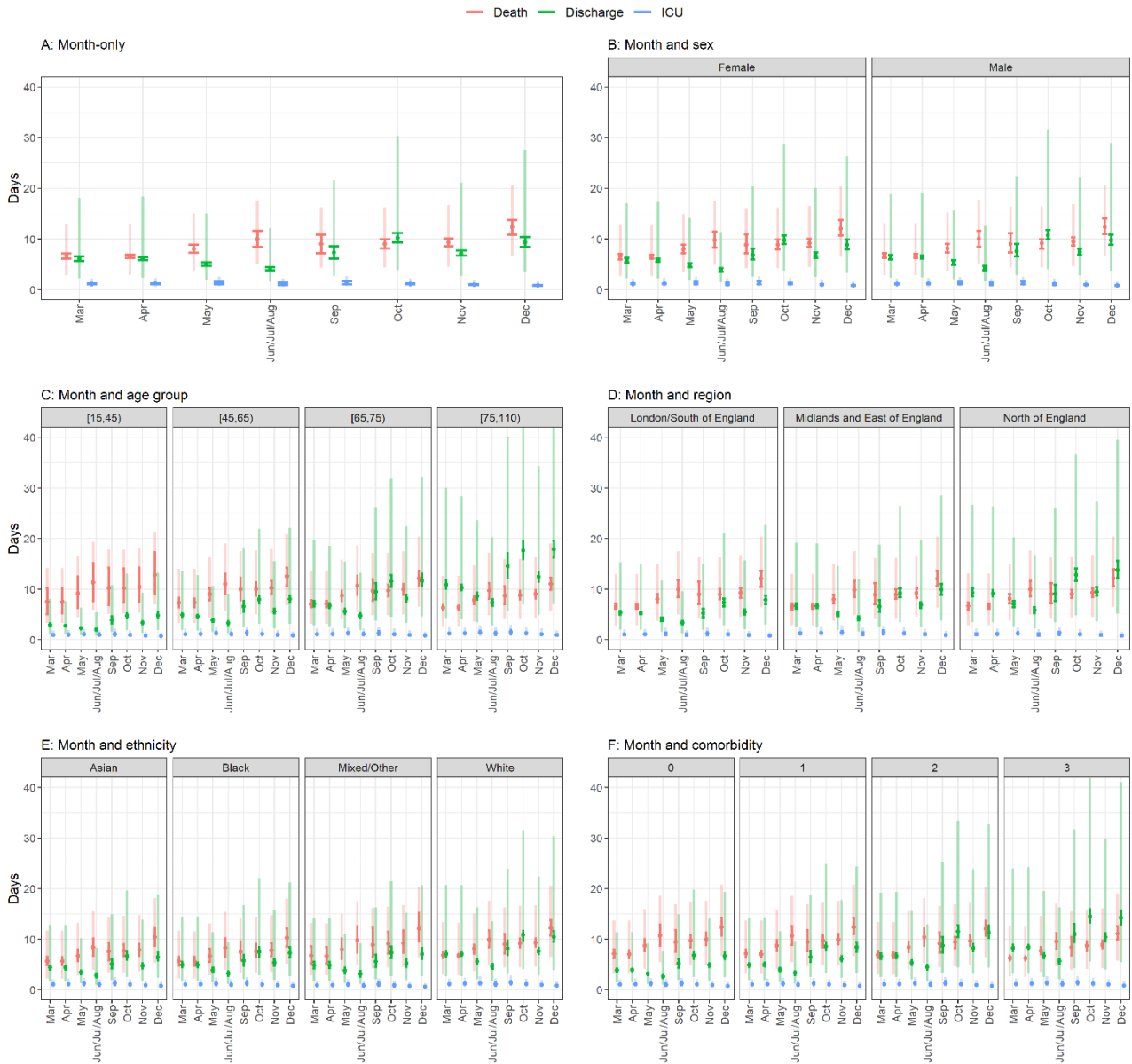

Figure 8: Estimated median time to event following ICU admission ('From ICU' sub-model), by month of admission (panel A), sex (panel B), age group (panel C), region of residence (panel D), ethnicity (panel E), and number of comorbidities (panel F). Line ranges are interquartile ranges, representing variability in estimated times to event, error bars are 95% confidence intervals, representing uncertainty in the estimated median time.

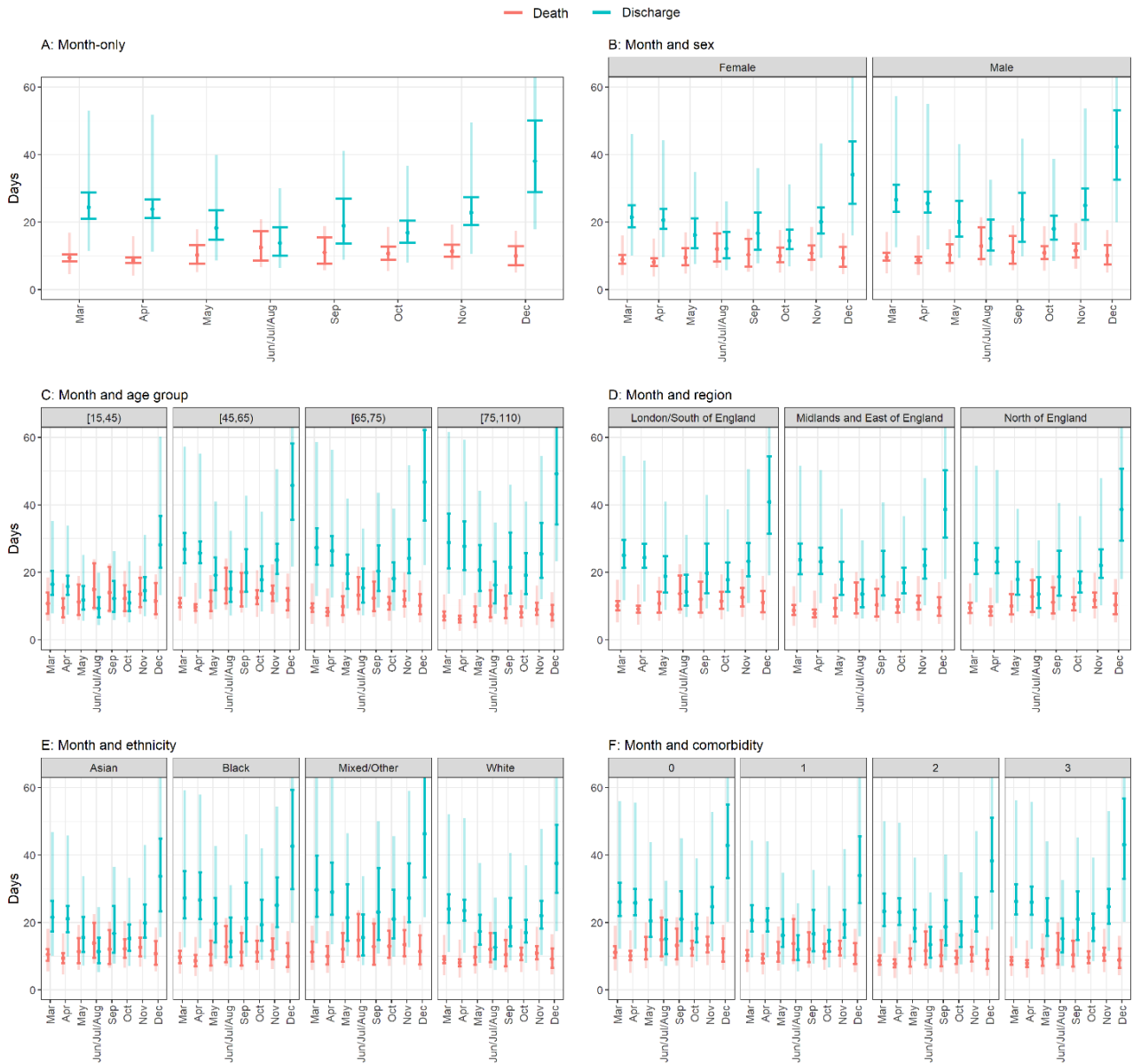

**Appendix**

**Model description**

We used a parametric semi-Markov multi-state model with four states. The model comprises two sub-models which represent competing risks of the next event 'from Hospital admission' and 'from ICU admission' (Figure 1).

In each of the sub-models an individual $i$ in state $r$ makes a transition to a state $s$ with probability $\pi_{r,s}$. Denote $S_{r,s}$ the distribution of time from entering state $r$ to moving to state $s$, given that this transition occurs (i.e. the conditional length of stay). A variety of parametric distributions are considered for this time.

This is a semi-Markov model, as the rates of transition to subsequent states depend on how long a person has spent in the current state, but not on any events before entering the current state.

*Censoring of data*

Transitions experienced by individual $i$ are observed in the data as one of three types $\delta_{i,j}$:

1. exact transition time: $\{y_{i,j}, r_{i,j}, s_{i,j}, \delta_{i,j} = 1\}$ transition from state $r_{i,j}$ to state $s_{i,j}$ occurs at time $y_{i,j}$

2. right censoring: $\{y_{i,j}, r_{i,j}, s_{i,j}, \delta_{i,j} = 2\}$ follow-up ends while individual is in state $r_{i,j}$ at time $y_{i,j}$, the next state and time of transition are unknown

3. partially-known outcome: $\{y_{i,j}, r_{i,j}, s_{i,j}, \delta_{i,j} = 3\}$ it is known whether an individual went to ICU, but unknown if they are still in hospital at time $y_{i,j}$

*Likelihood*

Let $f_{r,s}(\ |\theta_{r,s})$ be the density of the parametric distribution for the time of transition from state $r$ to state $s$, conditional on this being the transition that occurs. Then the likelihood contributions are:

$$\delta_{i,j} = 1: \qquad l_{i,j} = \pi_{r_{i,j}s_{i,j}} f_{r_{i,j}s_{i,j}}(y_{i,j}|\theta_{r_{i,j}s_{i,j}})$$

$$\delta_{i,j} = 2: \qquad l_{i,j} = \sum_{s \in S_r} \pi_{r_{i,j}s}(1 - F_{r_{i,j}s}(y_{i,j}|\theta_{r_{i,j}s}))$$

$$\delta_{i,j} = 3, s \neq \text{Discharge:} \qquad \text{as for } \delta_{i,j} = 2$$

$$\delta_{i,j} = 3, s = \text{Discharge:} \qquad l_{i,j} = \pi_{r_{i,j}s}$$

*Covariates*

The flexsurv package in R is used to maximise this likelihood for the SARI-Watch dataset. Multinomial logistic regression is used within the flexsurv package to allow the $\pi_{r,s}$ to depend upon the covariates and their interactions. Covariates are also included on parameters $\theta_{r,s}$ of the parametric distributions $f_{r,s}$. The choice of parameters and covariate effects are shown in Supplementary Table 1.

The parametric distributions chosen included the Generalized Gamma, Gamma and Log-normal distributions. The cumulative distribution functions for these distributions are included in [10].

**Data linkage**

Data linkage to PHE's mortality data was undertaken using a deterministic linkage on NHS number. PHE's mortality data comprised deaths from four sources in order to ensure maximum completeness (the Office for National Statistics (ONS) death registrations, NHSE deaths in hospitals, SGSS and local Health Protection Teams). Death outcomes are included in these data if they occurred within 60 days of a positive COVID-19 test, or COVID-19 was mentioned on the death certificate.

Data linkage to the NHS Hospital Episodes Statistics database was undertaken using a deterministic linkage on NHS number.

**Goodness of fit**

Goodness of fit curves for the parametric model compared to the non-parametric Aalen-Johansen cumulative incidence estimator are included in the Appendix. The models showed good agreement, with the parametric model estimating increased probability of discharge compared to the Aalen-Johansen estimator due to the two methods assuming different censoring mechanisms.

Supplementary table 1: Model parameters for 'From Hospital' and 'From ICU' sub-models

| | 'From Hospital' sub-model | | | 'From ICU' sub-model | |
|---|---|---|---|---|---|
| **Transition** | Discharge | Death | ICU | Discharge | Death |
| **Parametric distribution** | Generalised Gamma | Gamma | Log-normal | Generalised gamma | Gamma |
| **Covariates included on** | Location (μ) | Rate (β) | Mean (μ) | Location (μ) | Rate (β) |

Supplementary table 2: List of SARI-Watch sentinel NHS trusts:

| |
|---|
| Alder Hey Children's NHS Foundation Trust |
| Barking, Havering and Redbridge University Hospitals NHS Trust |
| Barts Health NHS Trust |
| Blackpool Teaching Hospitals NHS Foundation Trust |
| Bolton NHS Foundation Trust |
| Chesterfield Royal Hospital NHS Foundation Trust |
| Great Western Hospitals NHS Foundation Trust |
| James Paget University Hospitals NHS Foundation Trust |
| King's College Hospital NHS Foundation Trust |
| Liverpool Heart and Chest Hospital NHS Foundation Trust |
| Luton And Dunstable University Hospital NHS Foundation Trust |
| Mid Essex Hospital Services NHS Trust |
| North Cumbria Integrated Care NHS Foundation Trust |
| North West Anglia NHS Foundation Trust |
| Northampton General Hospital NHS Trust |
| Royal Berkshire NHS Foundation Trust |
| Royal National Orthopaedic Hospital NHS Trust |
| Shrewsbury And Telford Hospital NHS Trust |
| Tameside And Glossop Integrated Care NHS Foundation Trust |
| The Newcastle Upon Tyne Hospitals NHS Foundation Trust |
| The Queen Elizabeth Hospital, King's Lynn, NHS Foundation Trust |
| The Rotherham NHS Foundation Trust |
| The Royal Orthopaedic Hospital NHS Foundation Trust |
| University College London Hospitals NHS Foundation Trust |
| University Hospital Southampton NHS Foundation Trust |
| Walsall Healthcare NHS Trust |
| West Hertfordshire Hospitals NHS Trust |
| Worcestershire Acute Hospitals NHS Trust |
| Wrightington, Wigan and Leigh NHS Foundation Trust |
| Yeovil District Hospital NHS Foundation Trust |
| York Teaching Hospital NHS Foundation Trust |

Supplementary table 3: Estimated next event probabilities and 95% confidence intervals for outcomes from hospital admission, by baseline characteristic.

| Group | Event | Mar | Apr | May | Jun/Jul/Aug | Sep | Oct | Nov | Dec |
|---|---|---|---|---|---|---|---|---|---|
| All | Death | 25.3% (23.8, 26.8) | 24.4% (23.4, 25.4) | 18% (16.5, 19.7) | 9.4% (7.9, 11.1) | 16.9% (13.8, 20.1) | 18.7% (16.9, 20.5) | 19.3% (17.9, 21) | 23.1% (20.4, 26) |
| All | Discharge | 60.8% (59.1, 62.5) | 65% (63.9, 66.2) | 75.8% (73.9, 77.6) | 84.3% (82.1, 86.2) | 69% (64.6, 72.9) | 68.5% (66.4, 70.6) | 69.6% (67.7, 71.4) | 67.3% (64, 69.9) |
| All | ICU | 13.9% (12.8, 15.2) | 10.6% (10, 11.4) | 6.2% (5.3, 7.1) | 6.3% (5.1, 7.8) | 14.2% (11.1, 17.2) | 12.8% (11.3, 14.5) | 11.1% (10, 12.6) | 9.6% (8.6, 10.9) |
| **Sex** | | | | | | | | | |
| Female | Death | 24.1% (22.6, 25.5) | 23% (21.8, 24.3) | 16.8% (15.3, 18.3) | 8.6% (7.2, 10.1) | 16.2% (13.2, 19.5) | 17.8% (16, 19.6) | 18.3% (16.9, 19.9) | 21.6% (18.9, 24.5) |
| Female | Discharge | 66.2% (64.5, 67.7) | 69.6% (68.2, 71) | 78.9% (77, 80.7) | 87.1% (85.4, 88.8) | 74.1% (70.1, 77.7) | 73.2% (71.1, 75.2) | 73.8% (72.1, 75.5) | 71.6% (68.5, 74.5) |
| Female | ICU | 9.8% (8.8, 10.9) | 7.4% (6.8, 8.1) | 4.3% (3.6, 5.3) | 4.3% (3.4, 5.3) | 9.8% (7.4, 12.6) | 9.1% (7.9, 10.5) | 7.9% (6.9, 8.9) | 6.8% (5.9, 7.8) |
| Male | Death | 26% (24.4, 27.6) | 25.4% (24.3, 26.6) | 19.3% (17.7, 21.3) | 10.1% (8.6, 11.7) | 17.8% (14.4, 21.3) | 19.6% (17.6, 21.4) | 20.3% (18.8, 22.1) | 24.1% (21, 27.2) |
| Male | Discharge | 57.2% (55.4, 58.9) | 61.5% (60.2, 62.8) | 72.7% (70.5, 74.5) | 81.8% (79.5, 84.1) | 65.1% (60.1, 69.7) | 64.5% (62.1, 66.7) | 65.7% (63.7, 67.8) | 63.9% (60.5, 67.3) |
| Male | ICU | 16.8% (15.6, 18.2) | 13.1% (12.1, 14) | 8% (6.7, 9.6) | 8.1% (6.5, 9.9) | 17.1% (13.4, 21.3) | 15.9% (14.2, 18) | 14% (12.4, 15.5) | 12% (10.7, 13.6) |
| **Age group** | | | | | | | | | |
| 15-44 | Death | 1.8% (0.9, 3.5) | 1.9% (1.2, 3.2) | 0.6% (0.2, 2.7) | 0.7% (0.1, 3.1) | 0.8% (0.1, 6.3) | 0.8% (0.2, 3.4) | 0.7% (0.2, 2.4) | 1.6% (0.4, 4.9) |
| 15-44 | Discharge | 85% (80.9, 88.1) | 85.9% (82.9, 88.4) | 95.1% (91.5, 96.7) | 93.1% (88.7, 95.7) | 90.2% (82.6, 94.6) | 90.7% (86.1, 93.8) | 91.5% (88.4, 94.3) | 86.6% (81.6, 90.4) |
| 15-44 | ICU | 13.2% (10.4, 16.9) | 12.2% (9.7, 14.8) | 4.3% (2.6, 7.3) | 6.2% (3.8, 10) | 9% (4.6, 15.7) | 8.5% (5.5, 12.8) | 7.8% (5.3, 10.8) | 11.8% (8.5, 15.6) |
| 45-64 | Death | 7.3% (5.9, 8.9) | 6.9% (5.8, 8.2) | 5.8% (4.1, 7.8) | 1.9% (0.9, 3.8) | 4.1% (1.6, 9.1) | 4.7% (3.1, 6.8) | 5.9% (4.3, 8) | 4.5% (2.8, 7.3) |
| 45-64 | Discharge | 69.2% (66.5, 72) | 72.6% (70.6, 74.7) | 80.3% (76.4, 83.2) | 87.6% (83.9, 90.3) | 72.2% (63.7, 79) | 74.9% (71.1, 78.5) | 75.2% (71.5, 78.5) | 78.3% (74.7, 81.9) |
| 45-64 | ICU | 23.5% (20.9, 26.3) | 20.5% (18.7, 22.3) | 13.9% (11.5, 17.4) | 10.4% (7.9, 14) | 23.7% (17.4, 31.3) | 20.4% (17.1, 24.3) | 18.9% (16, 22.2) | 17.1% (14.2, 20.1) |
| 65-74 | Death | 21.7% (18.7, 25) | 22.8% (20.6, 25.2) | 16.4% (13.1, 20.5) | 7.9% (5, 12.1) | 17.4% (10.1, 27.9) | 17.6% (14.5, 21.2) | 17.3% (14.5, 20.8) | 19% (13.8, 24.4) |
| 65-74 | Discharge | 59.1% (55.3, 63) | 62.4% (60, 65.1) | 76% (71.2, 79.6) | 86.1% (80.7, 89.9) | 63.5% (52.3, 72.8) | 62.6% (57.8, 66.9) | 65% (60.7, 68.8) | 69.9% (63.7, 75.3) |
| 65-74 | ICU | 19.2% (16.3, 22.3) | 14.8% (12.8, 16.8) | 7.5% (5.3, 10.5) | 6% (3.5, 10) | 19.1% (12.1, 28.8) | 19.7% (16.1, 23.9) | 17.6% (14.5, 21) | 11.1% (8.4, 14.7) |
| 75+ | Death | 47.5% (44.7, 50.1) | 42% (40.4, 43.8) | 30.3% (27.4, 33.3) | 21.2% (17.7, 25) | 37.1% (29.8, 44.9) | 33.8% (30.6, 37.2) | 34.2% (31.3, 37) | 41% (36.6, 45.5) |
| 75+ | Discharge | 47.6% (45, 50.2) | 56% (54.2, 57.6) | 67.5% (64.5, 70.6) | 76.4% (72.1, 79.7) | 56% (48.8, 64.1) | 60.9% (57.6, 64.4) | 61.2% (58.4, 63.9) | 56.7% (52.1, 61.3) |
| 75+ | ICU | 4.9% (3.9, 6.2) | 2% (1.6, 2.6) | 2.2% (1.5, 3.4) | 2.4% (1.4, 4.2) | 6.9% (3.9, 12.4) | 5.3% (3.9, 7) | 4.6% (3.6, 6) | 2.4% (1.6, 3.6) |
| **Region of residence** | | | | | | | | | |
| London/South of England | Death | 22.9% (21, 24.7) | 19.3% (18.1, 20.6) | 12.7% (10.6, 14.9) | 3.7% (2.4, 5.6) | 15.8% (8.6, 27.7) | 19.2% (14.2, 25.9) | 14.6% (11.7, 18.3) | 17.1% (13.6, 20.7) |
| London/South of England | Discharge | 63.5% (61.6, 65.6) | 69.8% (68.5, 71.2) | 81.3% (78.6, 83.7) | 92.5% (89.5, 94.4) | 72.6% (57.8, 82.6) | 70.2% (62.6, 76.2) | 75.4% (71.6, 79.3) | 74.2% (70.3, 77.8) |
| London/South of England | ICU | 13.7% (12.2, 15.2) | 10.8% (9.8, 11.8) | 6.1% (4.7, 8) | 3.8% (2.4, 5.9) | 11.7% (5.4, 22.4) | 10.5% (7, 15.2) | 10% (7.5, 12.6) | 8.7% (6.8, 11.2) |
| Midlands and East of England | Death | 30.5% (27.5, 33.5) | 32.6% (30.8, 34.7) | 24.4% (21.3, 27.6) | 13.2% (10.3, 16.6) | 14.7% (9.4, 22.5) | 14.8% (12.1, 18) | 18.6% (16.2, 21) | 24.6% (20.7, 28.9) |
| Midlands and East of England | Discharge | 55.6% (52.3, 59) | 59.3% (57.1, 61.2) | 69.9% (66.7, 73.1) | 79% (74.6, 82.6) | 74.6% (66.4, 80.8) | 71.7% (68.1, 74.9) | 72.3% (69.4, 74.8) | 68% (63.5, 72.1) |
| Midlands and East of England | ICU | 13.9% (12, 16.6) | 8% (7, 9.3) | 5.7% (4.2, 7.5) | 7.8% (5.7, 10.6) | 10.7% (6.8, 17.3) | 13.5% (11, 16.6) | 9.1% (7.7, 11) | 7.4% (6, 9) |
| North of England | Death | 24% (20.8, 28.1) | 26.7% (24.4, 28.9) | 17.8% (15.4, 20.7) | 12% (9.6, 15.3) | 17.9% (13.7, 23.1) | 20.8% (18.6, 23.3) | 22.7% (20, 25.4) | 25.5% (20.1, 31.6) |
| North of England | Discharge | 58.8% (54.1, 62.8) | 59.5% (56.8, 62.1) | 75.7% (72.3, 78.5) | 81.3% (77.5, 84.4) | 65.3% (58.7, 70.6) | 66.2% (63.2, 68.9) | 63.7% (60.8, 66.8) | 59.6% (52.9, 65.5) |
| North of England | ICU | 17.3% (14, 20.8) | 13.8% (11.9, 15.7) | 6.6% (5.1, 8.5) | 6.7% (4.7, 9.5) | 16.7% (12.9, 21.6) | 13% (11.1, 15) | 13.6% (11.8, 15.7) | 14.9% (12, 18.1) |
| **Ethnicity** | | | | | | | | | |

| Group | Event | Mar | Apr | May | Jun/Jul/Aug | Sep | Oct | Nov | Dec |
|---|---|---|---|---|---|---|---|---|---|
| Asian | Death | 20.4% (16.8, 24.6) | 17.7% (15, 20.3) | 6.2% (3.4, 10.7) | 3.4% (1.5, 7.5) | 11.4% (5.9, 21.4) | 18.4% (12.5, 24.1) | 9.6% (6.2, 14.4) | 9.9% (5.5, 16.5) |
| Asian | Discharge | 60.8% (56, 65.2) | 64.7% (61.4, 67.8) | 86.9% (81.1, 90.9) | 88.4% (82.3, 92.5) | 73.3% (61.4, 81.3) | 64.4% (56.7, 72.4) | 73.5% (66.3, 79.3) | 79.2% (71.2, 85.2) |
| Asian | ICU | 18.7% (15.3, 22.5) | 17.7% (15.3, 20.2) | 6.9% (4.2, 11.7) | 8.2% (4.4, 13.3) | 15.3% (8.8, 24.8) | 17.2% (11.9, 24.6) | 16.9% (12.3, 23.6) | 10.9% (7.1, 16.9) |
| Black | Death | 20.1% (16.6, 23.7) | 14.3% (11.7, 17.1) | 4.4% (2, 9.3) | 0.7% (0.1, 7.3) | 12.4% (2.4, 43.9) | 15.1% (5.9, 34.3) | 26.2% (13.2, 45.1) | 14.2% (6.1, 27.7) |
| Black | Discharge | 64% (59.7, 68.2) | 72.8% (69.3, 76.1) | 83.4% (75.6, 88.7) | 95.8% (85.8, 98.3) | 61.3% (29.7, 82.3) | 73.4% (51.6, 86.1) | 60.1% (41, 75.3) | 77.6% (62.2, 87) |
| Black | ICU | 15.9% (12.9, 19.1) | 12.9% (10.5, 15.6) | 12.2% (7.7, 19.4) | 3.6% (1.1, 9.7) | 26.3% (7.9, 54.1) | 11.4% (3.5, 32.8) | 13.7% (5.3, 29.8) | 8.2% (3.4, 17) |
| Mixed/Other | Death | 13.2% (8.6, 19.1) | 11.3% (8.4, 15.6) | 9% (4.6, 17.5) | 0.6% (0, 16.5) | 8.8% (1.4, 34) | 9.9% (3.4, 25.9) | 4.9% (1.4, 15.4) | 4.8% (1.4, 14.7) |
| Mixed/Other | Discharge | 66.8% (59.3, 72.9) | 70.2% (65.2, 74.4) | 82.7% (72.9, 88.9) | 95.7% (76.2, 98.8) | 75.8% (46.5, 89.9) | 62.9% (45.5, 77.8) | 72.4% (58.4, 83.3) | 81.2% (69.1, 88.6) |
| Mixed/Other | ICU | 20% (14.6, 26.1) | 18.5% (14.9, 23.2) | 8.3% (3.9, 17.1) | 3.7% (0.8, 12.9) | 15.4% (3.9, 39.7) | 27.2% (13.3, 44) | 22.7% (13.1, 35.3) | 13.9% (8.1, 23.8) |
| White | Death | 28.9% (27.3, 31.2) | 28.6% (27.4, 30.1) | 21% (19.2, 23) | 11.8% (9.8, 14) | 19.1% (15.3, 23.6) | 19.5% (17.6, 21.5) | 20.9% (19.1, 22.7) | 25.4% (22.2, 28.7) |
| White | Discharge | 58.7% (56.3, 60.4) | 63% (61.6, 64.3) | 73.9% (71.7, 75.8) | 82% (79.3, 84.2) | 67.7% (62.5, 72.2) | 68% (65.6, 70.2) | 68.9% (66.8, 70.9) | 66.4% (63, 69.7) |
| White | ICU | 12.3% (11.1, 13.8) | 8.4% (7.7, 9.1) | 5.1% (4.2, 6.2) | 6.2% (4.8, 7.8) | 13.2% (10.2, 17.4) | 12.5% (10.9, 14.1) | 10.2% (8.9, 11.6) | 8.2% (7.1, 9.5) |
| **Number of comorbidities** | | | | | | | | | |
| 0 | Death | 9.6% (7.5, 12.1) | 11% (9.5, 12.7) | 7.8% (5.6, 10.8) | 3.3% (1.7, 5.9) | 5.9% (2.8, 12) | 7% (4.9, 9.6) | 7.4% (5.6, 10) | 10.2% (7.5, 13.6) |
| 0 | Discharge | 74.9% (71, 78) | 74.1% (71.8, 76.3) | 85.4% (81.4, 88.5) | 92% (87.9, 94.5) | 82.7% (75, 88.1) | 80.3% (76.3, 84) | 81.2% (77.9, 84) | 78.2% (74, 81.8) |
| 0 | ICU | 15.5% (12.8, 18.6) | 14.9% (13.2, 16.7) | 6.8% (4.8, 10.1) | 4.7% (2.7, 7.7) | 11.4% (6.8, 17.8) | 12.7% (10, 16.4) | 11.4% (9.1, 14.2) | 11.6% (9.2, 14.2) |
| 1 | Death | 16% (13.3, 18.8) | 18% (15.9, 20.1) | 11.7% (9.2, 15.2) | 4.8% (2.9, 8.3) | 14.4% (8.8, 22.4) | 11.9% (8.8, 15.7) | 13.9% (11.1, 17.3) | 18.3% (14.4, 23.8) |
| 1 | Discharge | 65.4% (61.7, 68.8) | 68.8% (66.2, 71) | 79.1% (74.3, 82.4) | 88.9% (84.4, 92.1) | 76.3% (67.5, 83.4) | 73.5% (68.2, 78.1) | 72.4% (68.3, 76.1) | 72% (66.6, 76.3) |
| 1 | ICU | 18.6% (15.6, 22) | 13.2% (11.7, 15.2) | 9.2% (6.8, 12.8) | 6.3% (3.9, 10) | 9.3% (5.2, 16.8) | 14.6% (11.1, 18.8) | 13.7% (10.9, 17) | 9.7% (7.5, 12.6) |
| 2 | Death | 22.8% (19.7, 26.2) | 27% (24.8, 29.2) | 17.6% (13.9, 21.6) | 13.9% (10.1, 19.1) | 21% (14.2, 31.2) | 19.1% (15.6, 23.8) | 20.9% (17.6, 24.8) | 26.9% (21.5, 32.9) |
| 2 | Discharge | 61.1% (57.4, 64.7) | 61.9% (59.3, 64.6) | 75% (71, 78.9) | 79.8% (73.8, 84) | 61% (49.7, 71) | 65.9% (60.6, 69.9) | 67.4% (62.9, 71.3) | 62.1% (56, 67.9) |
| 2 | ICU | 16.1% (13.4, 19.1) | 11.1% (9.6, 12.9) | 7.4% (5.3, 10.5) | 6.3% (3.7, 10) | 18% (10.9, 26.3) | 15% (11.4, 19.3) | 11.7% (9, 14.9) | 10.9% (8.1, 14.5) |
| 3+ | Death | 35.8% (33.6, 38.2) | 32.3% (30.8, 33.8) | 25.1% (22.4, 27.8) | 13.5% (10.8, 16.2) | 26.3% (19.4, 33.8) | 28.8% (25.7, 32.1) | 29.8% (27, 32.9) | 31.9% (27.4, 37) |
| 3+ | Discharge | 53% (50.4, 55.8) | 60.4% (58.6, 62) | 70.8% (67.9, 73.6) | 79.9% (75.9, 82.9) | 56.5% (48.2, 64.6) | 60.4% (57, 63.9) | 61.1% (57.8, 64.1) | 61.9% (56.8, 66.3) |
| 3+ | ICU | 11.2% (9.5, 12.9) | 7.3% (6.5, 8.4) | 4.1% (3, 5.4) | 6.7% (4.9, 9.2) | 17.2% (11.9, 24.8) | 10.8% (8.8, 13.1) | 9.2% (7.5, 11.2) | 6.3% (4.7, 8.8) |

Supplementary table 4: Estimated next event probabilities and 95% confidence intervals for outcome from ICU admission, by baseline characteristic.

| Group | Event | Mar | Apr | May | Jun/Jul/Aug | Sep | Oct | Nov | Dec |
|---|---|---|---|---|---|---|---|---|---|
| All | Death | 47.4% (42.5, 51.7) | 47.5% (44, 51.2) | 31% (23.1, 39.8) | 24.1% (16.1, 35.1) | 33.4% (23.5, 45.4) | 40.2% (34.1, 46.6) | 37% (31.7, 42.7) | 26.8% (20.2, 34.2) |
| All | Discharge | 52.6% (48.3, 57.5) | 52.5% (48.8, 56) | 69% (60.2, 76.9) | 75.9% (64.9, 83.9) | 66.6% (54.6, 76.5) | 59.8% (53.4, 65.9) | 63% (57.3, 68.3) | 73.2% (65.8, 79.8) |
| **Sex** | | | | | | | | | |
| Female | Death | 34.8% (26.9, 42.8) | 41.7% (35.2, 48.2) | 24.5% (14.7, 39.1) | 16.2% (7.2, 35.5) | 21.3% (10.2, 42.4) | 24.1% (15, 35.3) | 32.6% (24.7, 42.9) | 23.6% (13.5, 36.6) |
| Female | Discharge | 65.2% (57.2, 73.1) | 58.3% (51.8, 64.8) | 75.5% (60.9, 85.3) | 83.8% (64.5, 92.8) | 78.7% (57.6, 89.8) | 75.9% (64.7, 85) | 67.4% (57.1, 75.3) | 76.4% (63.4, 86.5) |
| Male | Death | 54.7% (49.1, 59.3) | 49.7% (45.2, 53.8) | 32.6% (24.4, 42.6) | 26.7% (16.7, 41.5) | 37.4% (24, 51.7) | 46.7% (39.4, 53.4) | 39.8% (31.9, 48) | 26.5% (19.2, 36) |
| Male | Discharge | 45.3% (40.7, 50.9) | 50.3% (46.2, 54.8) | 67.4% (57.4, 75.6) | 73.3% (58.5, 83.3) | 62.6% (48.3, 76) | 53.3% (46.6, 60.6) | 60.2% (52, 68.1) | 73.5% (64, 80.8) |
| **Age group** | | | | | | | | | |
| 15-44 | Death | 20.5% (11.6, 32.4) | 13% (7.5, 21.7) | 10.1% (2, 43.7) | 8.2% (1.3, 36.9) | 11.1% (0.8, 66) | 13.1% (3.7, 35.8) | 10.4% (3.2, 27.6) | 9.3% (3, 26.3) |
| 15-44 | Discharge | 79.5% (67.6, 88.4) | 87% (78.3, 92.5) | 89.9% (56.3, 98) | 91.8% (63.1, 98.7) | 88.9% (34, 99.2) | 86.9% (64.2, 96.3) | 89.6% (72.4, 96.8) | 90.7% (73.7, 97) |
| 45-64 | Death | 42.2% (36, 47.6) | 43.7% (39, 48.5) | 24.3% (16.5, 35.8) | 14% (6.1, 27.5) | 10% (3.6, 27.5) | 24% (16.2, 34.7) | 27.5% (20.2, 35.9) | 23.7% (15.4, 36.7) |
| 45-64 | Discharge | 57.8% (52.4, 64) | 56.3% (51.5, 61) | 75.7% (64.2, 83.5) | 86% (72.5, 93.9) | 90% (72.5, 96.4) | 76% (65.3, 83.8) | 72.5% (64.1, 79.8) | 76.3% (63.3, 84.6) |

| Group | Type | Mar | Apr | May | Jun/Jul/Aug | Sep | Oct | Nov | Dec |
|---|---|---|---|---|---|---|---|---|---|
| 65-74 | Death | 63.1% (55, 71.5) | 61.4% (53.7, 68.2) | 33.6% (17.7, 53.5) | 39.9% (18.3, 65.3) | 59.2% (32.8, 79.2) | 49.3% (39.1, 61.2) | 40.9% (30.8, 51) | 27.3% (14, 46.1) |
| 65-74 | Discharge | 36.9% (28.5, 45) | 38.6% (31.8, 46.3) | 66.4% (46.5, 82.3) | 60.1% (34.7, 81.7) | 40.8% (20.8, 67.2) | 50.7% (38.8, 60.9) | 59.1% (49, 69.2) | 72.7% (53.9, 86) |
| 75+ | Death | 77.4% (65.1, 85.9) | 75.3% (65, 84.8) | 51.9% (31.2, 71.5) | 49.5% (25.6, 75) | 60.2% (32.6, 82) | 76% (61.3, 86.1) | 65.1% (49.2, 77.3) | 70.5% (44.4, 88.5) |
| 75+ | Discharge | 22.6% (14.1, 34.9) | 24.7% (15.2, 35) | 48.1% (28.5, 68.8) | 50.5% (25, 74.4) | 39.8% (18, 67.4) | 24% (13.9, 38.7) | 34.9% (22.7, 50.8) | 29.5% (11.5, 55.6) |
| **Region of residence** | | | | | | | | | |
| London/South of England | Death | 47.1% (41.1, 53.1) | 48.3% (43.5, 52.5) | 30.4% (20, 44.5) | 23.8% (9.3, 46.8) | 32.5% (5.7, 74.5) | 44.4% (21.1, 69.6) | 24.3% (14.3, 38.6) | 24.8% (14.1, 39.6) |
| London/South of England | Discharge | 52.9% (46.9, 58.9) | 51.7% (47.5, 56.5) | 69.6% (55.5, 80) | 76.2% (53.2, 90.7) | 67.5% (25.5, 94.3) | 55.6% (30.4, 78.9) | 75.7% (61.4, 85.7) | 75.2% (60.4, 85.9) |
| Midlands and East of England | Death | 57% (47.9, 64.9) | 57.9% (50.4, 65.9) | 38.3% (24.7, 52.6) | 21% (11.3, 37.8) | 35.6% (15.4, 61.8) | 50.6% (40.7, 60.7) | 47.1% (37.4, 56.8) | 32.5% (21.7, 44) |
| Midlands and East of England | Discharge | 43% (35.1, 52.1) | 42.1% (34.1, 49.6) | 61.7% (47.4, 75.3) | 79% (62.2, 88.7) | 64.4% (38.2, 84.6) | 49.4% (39.3, 59.3) | 52.9% (43.2, 62.6) | 67.5% (56, 78.3) |
| North of England | Death | 47.3% (36.9, 57.2) | 35.6% (28.6, 43) | 19.5% (10.8, 34.2) | 21.9% (9.5, 40.8) | 30.2% (18.5, 44.7) | 33.8% (27.3, 40.9) | 33.7% (26.1, 42.4) | 21.1% (12.4, 34.8) |
| North of England | Discharge | 52.7% (42.8, 63.1) | 64.4% (57, 71.4) | 80.5% (65.8, 89.2) | 78.1% (59.2, 90.5) | 69.8% (55.3, 81.5) | 66.2% (59.1, 72.7) | 66.3% (57.6, 73.9) | 78.9% (65.2, 87.6) |
| **Ethnicity** | | | | | | | | | |
| Asian | Death | 52.2% (45.8, 58.9) | 51.8% (46.1, 58) | 33.6% (24, 43.8) | 25.7% (16.2, 37.7) | 35% (22.9, 48.1) | 44.7% (36.1, 52.5) | 39.5% (31.9, 48.3) | 28.3% (19.9, 37.4) |
| Asian | Discharge | 47.8% (41.1, 54.2) | 48.2% (42, 53.9) | 66.4% (56.2, 76) | 74.3% (62.3, 83.8) | 65% (51.9, 77.1) | 55.3% (47.5, 63.9) | 60.5% (51.7, 68.1) | 71.7% (62.6, 80.1) |
| Black | Death | 48.2% (41.1, 55.7) | 47.8% (40, 55.2) | 30.2% (21.7, 41.1) | 22.8% (13.5, 34) | 31.5% (20, 44.5) | 40.8% (31.9, 51) | 35.8% (27.2, 45.3) | 25.2% (16.8, 35.4) |
| Black | Discharge | 51.8% (44.3, 58.9) | 52.2% (44.8, 60) | 69.8% (58.9, 78.3) | 77.2% (66, 86.5) | 68.5% (55.5, 80) | 59.2% (49, 68.1) | 64.2% (54.7, 72.8) | 74.8% (64.6, 83.2) |
| Mixed/Other | Death | 37.6% (29.1, 46.7) | 37.3% (28.9, 45.9) | 21.9% (13.7, 31.5) | 16.1% (8.6, 27) | 23% (13.3, 34.2) | 30.9% (22.3, 41.7) | 26.5% (18.3, 36.2) | 17.9% (11.5, 26.8) |
| Mixed/Other | Discharge | 62.4% (53.3, 70.9) | 62.7% (54.1, 71.1) | 78.1% (68.5, 86.3) | 83.9% (73, 91.4) | 77% (65.8, 86.7) | 69.1% (58.3, 77.7) | 73.5% (63.8, 81.7) | 82.1% (73.2, 88.5) |
| White | Death | 48.5% (43.4, 53.3) | 48.1% (44.1, 52.4) | 30.4% (22.6, 38.6) | 23% (14.6, 33.7) | 31.7% (21, 43.1) | 41% (34.8, 47.4) | 36% (29.4, 42.5) | 25.4% (18.4, 32.4) |
| White | Discharge | 51.5% (46.7, 56.6) | 51.9% (47.6, 55.9) | 69.6% (61.4, 77.4) | 77% (66.3, 85.4) | 68.3% (56.9, 79) | 59% (52.6, 65.2) | 64% (57.5, 70.6) | 74.6% (67.6, 81.6) |
| **Number of comorbidities** | | | | | | | | | |
| 0 | Death | 42.2% (32.2, 51.6) | 40.1% (32.8, 46.6) | 24.4% (10.7, 45.4) | 21.3% (6.5, 51.1) | 25% (10.1, 51.9) | 41.9% (28.7, 54.4) | 34% (23.5, 45.3) | 22.1% (12.4, 35.6) |
| 0 | Discharge | 57.8% (48.4, 67.8) | 59.9% (53.4, 67.2) | 75.6% (54.6, 89.3) | 78.7% (48.9, 93.5) | 75% (48.1, 89.9) | 58.1% (45.6, 71.3) | 66% (54.7, 76.5) | 77.9% (64.4, 87.6) |
| 1 | Death | 43.4% (33.7, 52.6) | 47.5% (40.7, 55.4) | 32.4% (19.2, 49.4) | 15.9% (5.6, 39.9) | 28.2% (10.5, 57.5) | 36.2% (24.9, 49.3) | 36.1% (25.7, 47.6) | 25.1% (12.3, 42.8) |
| 1 | Discharge | 56.6% (47.4, 66.3) | 52.5% (44.6, 59.3) | 67.6% (50.6, 80.8) | 84.1% (60.1, 94.4) | 71.8% (42.5, 89.5) | 63.8% (50.7, 75.1) | 63.9% (52.4, 74.3) | 74.9% (57.2, 87.7) |
| 2 | Death | 57.4% (47.4, 65.8) | 48.9% (40.9, 56.4) | 29.5% (15.9, 48.2) | 12.6% (3.6, 39) | 33.5% (16.3, 60.8) | 41.7% (30.3, 53.6) | 42.2% (29.7, 54.9) | 16.6% (7.3, 32.4) |
| 2 | Discharge | 42.6% (34.2, 52.6) | 51.1% (43.6, 59.1) | 70.5% (51.8, 84.1) | 87.4% (61, 96.4) | 66.5% (39.2, 83.7) | 58.3% (46.4, 69.7) | 57.8% (45.1, 70.3) | 83.4% (67.6, 92.7) |
| 3+ | Death | 50.7% (43.2, 58.3) | 53% (46.6, 59.2) | 33.5% (22.3, 46.2) | 28.2% (15.5, 44.8) | 30.4% (15.1, 51.8) | 42% (31.4, 52.3) | 38.6% (28.2, 48.9) | 42% (27, 57.4) |
| 3+ | Discharge | 49.3% (41.7, 56.8) | 47% (40.8, 53.4) | 66.5% (53.8, 77.7) | 71.8% (55.2, 84.5) | 69.6% (48.2, 84.9) | 58% (47.7, 68.6) | 61.4% (51.1, 71.8) | 58% (42.6, 73) |

Supplementary table 5: Estimated hospitalised fatality risk and 95% confidence intervals, by baseline characteristic.

| Group | Mar | Apr | May | Jun/Jul/Aug | Sep | Oct | Nov | Dec |
|---|---|---|---|---|---|---|---|---|
| All | 31.9% (30.3, 33.5) | 29.5% (28.4, 30.5) | 19.9% (18.3, 21.6) | 10.9% (9.4, 12.7) | 21.6% (18.4, 25.5) | 23.8% (22.1, 25.9) | 23.4% (21.8, 25) | 25.7% (23, 29.2) |
| **Sex** | | | | | | | | |
| Female | 27.5% (25.6, 29.4) | 26.1% (24.8, 27.4) | 17.9% (16.4, 19.7) | 9.3% (7.9, 11.1) | 18.2% (14.8, 22.2) | 20% (18.2, 21.9) | 20.9% (19.3, 22.6) | 23.2% (20.3, 26.3) |

| | Mar | Apr | May | Jun/Jul/Aug | Sep | Oct | Nov | Dec |
|---|---|---|---|---|---|---|---|---|
| Male | 35.2% (33.4, 36.9) | 31.9% (30.6, 33.2) | 21.9% (20.2, 23.9) | 12.2% (10.4, 14.3) | 24.2% (20.2, 28.5) | 27% (24.9, 29.1) | 25.9% (24.1, 28) | 27.3% (24.2, 30.4) |
| **Age group** | | | | | | | | |
| 15-44 | 4.5% (2.9, 6.8) | 3.5% (2.4, 5.1) | 1.1% (0.4, 3.5) | 1.2% (0.4, 4.5) | 1.8% (0.3, 9.5) | 1.9% (0.8, 5.3) | 1.5% (0.6, 3.8) | 2.6% (1.3, 6.9) |
| 45-64 | 17.2% (15.1, 19.6) | 15.9% (14.4, 17.5) | 9.2% (6.9, 11.9) | 3.4% (2, 5.7) | 6.4% (3.6, 12.2) | 9.6% (7.3, 12.4) | 11.1% (8.8, 13.9) | 8.6% (6.3, 11.9) |
| 65-74 | 33.8% (30, 37.8) | 31.9% (29.6, 34.6) | 19% (15.2, 23.6) | 10.3% (7, 15) | 28.7% (19.3, 39.9) | 27.4% (23.4, 32.1) | 24.6% (21.4, 28.3) | 22% (16.6, 27.3) |
| 75+ | 51.3% (48.6, 54.1) | 43.5% (41.7, 45.3) | 31.4% (28.8, 34.1) | 22.4% (18.8, 26.4) | 41.2% (34.1, 49) | 37.8% (34.4, 41.4) | 37.2% (34.1, 40) | 42.6% (38.3, 46.8) |
| **Region of residence** | | | | | | | | |
| London/South of England | 29.3% (27.4, 31.2) | 24.6% (23.1, 26) | 14.5% (12.4, 17.1) | 4.6% (3.1, 7.1) | 19.6% (11.2, 33.1) | 23.9% (18, 30.2) | 17% (14.3, 20.4) | 19.3% (16.1, 23.3) |
| Midlands and East of England | 38.4% (35.2, 41.9) | 37.3% (35.1, 39.4) | 26.5% (23.5, 29.6) | 14.9% (12, 18.5) | 18.5% (13, 26.8) | 21.6% (18.7, 25.1) | 22.9% (20.6, 25.8) | 27% (22.9, 31.6) |
| North of England | 32.1% (27.9, 36.7) | 31.6% (29.3, 34.1) | 19.1% (16.2, 22.3) | 13.5% (10.3, 17) | 23% (18.3, 29) | 25.2% (22.7, 28) | 27.3% (24.4, 30.2) | 28.6% (23.3, 34.8) |
| **Ethnicity** | | | | | | | | |
| Asian | 30.2% (26.3, 34.4) | 26.8% (24.2, 29.6) | 8.5% (5.6, 12.9) | 5.5% (3.3, 10) | 16.7% (10.6, 26.2) | 26% (20.2, 32.8) | 16.3% (12.6, 21.8) | 13% (9, 19.3) |
| Black | 27.8% (24.4, 31.8) | 20.5% (17.5, 23.7) | 8.1% (4.9, 14.1) | 1.5% (0.5, 9.7) | 20.7% (7.9, 48) | 19.8% (10.1, 37.7) | 31.1% (18, 47.6) | 16.2% (8.2, 32.5) |
| Mixed/Other | 20.7% (15.8, 27.2) | 18.2% (14.9, 22.4) | 10.8% (6.1, 18.5) | 1.2% (0.3, 17.2) | 12.3% (4.2, 36.3) | 18.3% (9.8, 35) | 10.9% (6.5, 22) | 7.3% (3.3, 18.6) |
| White | 34.9% (33, 37) | 32.6% (31.4, 34) | 22.6% (20.8, 24.8) | 13.2% (11.5, 15.7) | 23.3% (19.4, 28.1) | 24.6% (22.6, 26.8) | 24.6% (22.8, 26.3) | 27.4% (24.3, 30.8) |
| **Number of comorbidities** | | | | | | | | |
| 0 | 16.1% (13.4, 19.1) | 17% (15.2, 18.9) | 9.5% (7.1, 12.8) | 4.3% (2.5, 7.7) | 8.8% (5.1, 16.6) | 12.3% (9.3, 15.8) | 11.3% (9.1, 14.3) | 12.8% (9.6, 16.4) |
| 1 | 24.1% (20.8, 27.4) | 24.3% (21.9, 26.7) | 14.7% (11.6, 18) | 5.8% (3.5, 9.7) | 17% (11.9, 25.7) | 17.2% (13.7, 21.8) | 18.9% (15.6, 22.1) | 20.7% (16.2, 25.8) |
| 2 | 32% (28.9, 36.2) | 32.5% (29.9, 34.9) | 19.8% (16.2, 23.8) | 14.7% (10.7, 20.4) | 27% (18.6, 39.2) | 25.3% (21.4, 29.5) | 25.8% (22.4, 30.2) | 28.7% (23.5, 35.4) |
| 3+ | 41.4% (39, 43.8) | 36.2% (34.4, 37.7) | 26.5% (23.9, 29.4) | 15.3% (12.6, 18.7) | 31.5% (23.7, 39.6) | 33.4% (30.1, 37.1) | 33.3% (29.9, 36.5) | 34.5% (29.9, 38.9) |

Supplementary table 6: Estimated median time (in days) to next event from hospital admission and 95% confidence interval, by baseline characteristic.

| Group | Event | Mar | Apr | May | Jun/Jul/Aug | Sep | Oct | Nov | Dec |
|---|---|---|---|---|---|---|---|---|---|
| All | Death | 6.6 (6.2, 7.1) | 6.6 (6.3, 6.9) | 8 (7.4, 8.8) | 9.9 (8.6, 11.7) | 9 (7.2, 11.1) | 9 (8.2, 9.9) | 9.3 (8.5, 10.1) | 12.3 (10.9, 14) |
| All | Discharge | 6.1 (5.7, 6.5) | 6.2 (5.9, 6.5) | 5.1 (4.8, 5.4) | 4.1 (3.8, 4.5) | 7.3 (6.2, 8.5) | 10.2 (9.3, 11.2) | 7.1 (6.7, 7.7) | 9.3 (8.5, 10.3) |
| All | ICU | 1.1 (1, 1.2) | 1.2 (1.1, 1.2) | 1.3 (1.1, 1.5) | 1.1 (0.9, 1.4) | 1.3 (1, 1.7) | 1.1 (1, 1.3) | 1 (0.9, 1.1) | 0.8 (0.7, 0.9) |
| **Sex** | | | | | | | | | |
| Female | Death | 6.5 (6, 7) | 6.5 (6.1, 6.9) | 7.9 (7.1, 8.8) | 9.8 (8.3, 11.5) | 8.8 (7.1, 11) | 8.9 (8, 9.9) | 9.2 (8.4, 10) | 12 (10.6, 13.6) |
| Female | Discharge | 5.8 (5.4, 6.2) | 5.9 (5.5, 6.2) | 4.8 (4.4, 5.2) | 3.9 (3.5, 4.3) | 6.9 (5.9, 8.2) | 9.8 (8.9, 10.6) | 6.8 (6.3, 7.3) | 8.9 (8, 10) |
| Female | ICU | 1.1 (1, 1.3) | 1.2 (1.1, 1.3) | 1.3 (1.1, 1.6) | 1.1 (0.9, 1.4) | 1.3 (1.1, 1.7) | 1.2 (1, 1.3) | 1 (0.9, 1.1) | 0.8 (0.7, 1) |
| Male | Death | 6.7 (6.2, 7.2) | 6.7 (6.3, 7.1) | 8.1 (7.3, 9) | 10 (8.6, 11.6) | 9 (7.4, 11.2) | 9.1 (8.3, 10.1) | 9.4 (8.6, 10.3) | 12.3 (10.9, 14) |
| Male | Discharge | 6.4 (6, 6.8) | 6.4 (6.1, 6.8) | 5.3 (4.9, 5.7) | 4.2 (3.8, 4.7) | 7.6 (6.5, 9) | 10.7 (9.8, 11.7) | 7.5 (6.9, 8.1) | 9.8 (8.8, 10.9) |
| Male | ICU | 1.1 (1, 1.2) | 1.2 (1.1, 1.2) | 1.3 (1.1, 1.5) | 1.1 (0.9, 1.3) | 1.3 (1, 1.7) | 1.1 (1, 1.3) | 1 (0.9, 1.1) | 0.8 (0.7, 0.9) |
| **Age group** | | | | | | | | | |
| 15-44 | Death | 7.5 (5.1, 10.5) | 7.5 (5.2, 10.4) | 9.2 (6.4, 12.9) | 11.3 (7.9, 15.9) | 10.2 (6.9, 14.4) | 10.2 (7.1, 14.3) | 10.4 (7.5, 14.7) | 12.8 (9, 18) |
| 15-44 | Discharge | 2.9 (2.7, 3.2) | 2.8 (2.6, 3) | 2.3 (2.1, 2.5) | 2 (1.8, 2.2) | 3.9 (3.3, 4.6) | 4.8 (4.3, 5.2) | 3.4 (3.1, 3.7) | 4.8 (4.4, 5.3) |
| 15-44 | ICU | 0.9 (0.8, 1.1) | 1 (0.9, 1.1) | 1.1 (0.9, 1.3) | 0.9 (0.7, 1.2) | 1.1 (0.9, 1.4) | 1 (0.8, 1.1) | 0.8 (0.7, 1) | 0.7 (0.6, 0.8) |
| 45-64 | Death | 7.3 (6.5, 8.2) | 7.3 (6.5, 8.2) | 9 (7.9, 10.3) | 11.1 (9.2, 13.1) | 10 (7.8, 12.3) | 10 (8.7, 11.4) | 10.2 (9, 11.5) | 12.5 (10.8, 14.6) |
| 45-64 | Discharge | 4.9 (4.6, 5.3) | 4.6 (4.4, 4.9) | 3.9 (3.6, 4.2) | 3.3 (3, 3.6) | 6.6 (5.6, 7.7) | 7.9 (7.2, 8.7) | 5.6 (5.2, 6.1) | 8 (7.3, 8.8) |
| 45-64 | ICU | 1.1 (1, 1.2) | 1.2 (1.1, 1.3) | 1.3 (1.1, 1.5) | 1.1 (0.9, 1.4) | 1.3 (1, 1.7) | 1.1 (1, 1.3) | 1 (0.9, 1.1) | 0.8 (0.7, 1) |
| 65-74 | Death | 7.1 (6.4, 7.8) | 7.1 (6.5, 7.7) | 8.7 (7.8, 9.7) | 10.7 (9.2, 12.4) | 9.7 (7.7, 11.8) | 9.7 (8.7, 10.9) | 9.9 (8.9, 10.9) | 12.2 (10.9, 13.8) |
| 65-74 | Discharge | 7.1 (6.6, 7.7) | 6.7 (6.3, 7.2) | 5.6 (5.1, 6.1) | 4.8 (4.4, 5.3) | 9.5 (8.2, 11.3) | 11.5 (10.5, 12.8) | 8.1 (7.5, 8.9) | 11.7 (10.5, 13) |
| 65-74 | ICU | 1.1 (1, 1.2) | 1.2 (1.1, 1.3) | 1.3 (1.1, 1.5) | 1.1 (0.9, 1.4) | 1.3 (1, 1.7) | 1.1 (1, 1.3) | 1 (0.9, 1.1) | 0.8 (0.7, 1) |
| 75+ | Death | 6.4 (5.9, 6.9) | 6.4 (6, 6.7) | 7.9 (7.1, 8.6) | 9.7 (8.4, 11.2) | 8.7 (7.2, 10.6) | 8.8 (7.9, 9.6) | 9 (8.3, 9.7) | 11.1 (10, 12.4) |
| 75+ | Discharge | 10.9 (10.1, 11.6) | 10.3 (9.8, 10.9) | 8.6 (7.9, 9.3) | 7.4 (6.7, 8.1) | 14.5 (12.4, 17.2) | 17.6 (16.1, 19.3) | 12.5 (11.5, 13.4) | 17.8 (16.2, 19.7) |
| 75+ | ICU | 1.3 (1.1, 1.4) | 1.3 (1.1, 1.5) | 1.5 (1.2, 1.8) | 1.3 (1, 1.6) | 1.5 (1.2, 2) | 1.3 (1.1, 1.5) | 1.1 (1, 1.3) | 0.9 (0.8, 1.1) |
| **Region of residence** | | | | | | | | | |
| London/South of England | Death | 6.6 (6.1, 7.1) | 6.6 (6.2, 7) | 8.1 (7.2, 9.1) | 9.9 (8.2, 11.5) | 8.9 (7.1, 11.1) | 9 (8, 10.1) | 9.3 (8.4, 10.2) | 12.1 (10.4, 13.7) |
| London/South of England | Discharge | 5.3 (5, 5.7) | 5.3 (5, 5.5) | 4.1 (3.7, 4.4) | 3.4 (3.1, 3.7) | 5.2 (4.4, 6.2) | 7.4 (6.7, 8.2) | 5.5 (5, 6) | 7.9 (7.2, 8.9) |
| London/South of England | ICU | 1.1 (1, 1.2) | 1.1 (1, 1.2) | 1.2 (1, 1.4) | 1 (0.8, 1.2) | 1.2 (0.9, 1.6) | 1 (0.9, 1.2) | 0.9 (0.8, 1) | 0.7 (0.6, 0.9) |
| Midlands and East of England | Death | 6.6 (6.1, 7.2) | 6.6 (6.2, 7) | 8 (7.3, 9) | 9.9 (8.4, 11.5) | 8.9 (7.2, 11) | 8.9 (8, 10) | 9.2 (8.5, 10.2) | 12 (10.6, 13.5) |
| Midlands and East of England | Discharge | 6.7 (6.2, 7.2) | 6.6 (6.2, 7) | 5.1 (4.7, 5.4) | 4.2 (3.8, 4.6) | 6.6 (5.5, 7.8) | 9.2 (8.4, 10.3) | 6.9 (6.3, 7.4) | 9.9 (9, 11.1) |
| Midlands and East of England | ICU | 1.3 (1.1, 1.4) | 1.3 (1.2, 1.5) | 1.5 (1.2, 1.7) | 1.2 (0.9, 1.5) | 1.5 (1.2, 1.9) | 1.3 (1.1, 1.5) | 1.1 (1, 1.3) | 0.9 (0.8, 1) |
| North of England | Death | 6.6 (6, 7.3) | 6.6 (6.1, 7.1) | 8.1 (7.3, 9) | 9.9 (8.5, 11.5) | 9 (7.3, 11.2) | 9 (8.1, 10) | 9.3 (8.5, 10.2) | 12.1 (10.5, 13.7) |
| North of England | Discharge | 9.3 (8.5, 10.1) | 9.2 (8.6, 9.8) | 7.1 (6.4, 7.7) | 5.8 (5.3, 6.5) | 9.1 (7.7, 10.6) | 12.8 (11.8, 14) | 9.5 (8.6, 10.4) | 13.8 (12.3, 15.6) |
| North of England | ICU | 1.1 (1, 1.3) | 1.2 (1, 1.3) | 1.3 (1.1, 1.5) | 1 (0.8, 1.3) | 1.3 (1, 1.6) | 1.1 (1, 1.2) | 0.9 (0.8, 1.1) | 0.8 (0.7, 0.9) |
| **Ethnicity** | | | | | | | | | |
| Asian | Death | 5.7 (5, 6.5) | 5.7 (5, 6.5) | 6.8 (5.8, 8) | 8.4 (6.9, 10.4) | 7.6 (6, 9.8) | 7.8 (6.6, 9) | 7.9 (6.8, 9.1) | 10.4 (8.7, 12.4) |
| Asian | Discharge | 4.4 (4, 4.8) | 4.4 (4.1, 4.8) | 3.5 (3.2, 3.9) | 2.9 (2.6, 3.2) | 5.1 (4.3, 6.2) | 6.7 (6, 7.5) | 4.8 (4.3, 5.3) | 6.5 (5.8, 7.3) |
| Asian | ICU | 1.1 (0.9, 1.2) | 1.1 (1, 1.3) | 1.3 (1, 1.5) | 1.1 (0.8, 1.3) | 1.3 (1, 1.7) | 1.1 (0.9, 1.3) | 0.9 (0.8, 1.1) | 0.8 (0.7, 1) |
| Black | Death | 5.7 (4.7, 6.5) | 5.6 (4.7, 6.5) | 6.7 (5.5, 7.9) | 8.4 (6.6, 10.4) | 7.5 (5.8, 9.6) | 7.7 (6.3, 9) | 7.8 (6.6, 9.1) | 10.3 (8.6, 12.3) |
| Black | Discharge | 5 (4.5, 5.5) | 5 (4.5, 5.4) | 3.9 (3.5, 4.4) | 3.3 (2.9, 3.6) | 5.8 (4.7, 7) | 7.6 (6.6, 8.6) | 5.4 (4.8, 6) | 7.3 (6.3, 8.3) |
| Black | ICU | 1.1 (0.9, 1.3) | 1.1 (1, 1.3) | 1.2 (1, 1.6) | 1 (0.8, 1.4) | 1.3 (1, 1.8) | 1.1 (0.9, 1.3) | 0.9 (0.8, 1.1) | 0.8 (0.7, 1) |
| Mixed/Other | Death | 6.8 (5.2, 8.6) | 6.7 (5.2, 8.4) | 8 (6.2, 10.2) | 9.9 (7.4, 12.9) | 8.9 (6.5, 12) | 9.1 (6.9, 11.4) | 9.3 (7.1, 11.6) | 12.1 (9.3, 15.5) |
| Mixed/Other | Discharge | 4.9 (4.3, 5.6) | 4.9 (4.3, 5.6) | 3.8 (3.3, 4.4) | 3.2 (2.8, 3.6) | 5.6 (4.6, 7) | 7.4 (6.3, 8.6) | 5.2 (4.6, 6.1) | 7.1 (6.1, 8.3) |
| Mixed/Other | ICU | 0.9 (0.8, 1.1) | 1 (0.8, 1.1) | 1.1 (0.8, 1.3) | 0.9 (0.7, 1.2) | 1.1 (0.8, 1.5) | 0.9 (0.7, 1.1) | 0.8 (0.7, 1) | 0.7 (0.6, 0.8) |
| White | Death | 6.9 (6.4, 7.4) | 6.8 (6.5, 7.2) | 8.1 (7.4, 8.9) | 10 (8.5, 11.7) | 9 (7.3, 11) | 9.2 (8.2, 10.1) | 9.4 (8.6, 10.1) | 12.2 (10.7, 13.8) |
| White | Discharge | 7.1 (6.6, 7.6) | 7.1 (6.8, 7.5) | 5.6 (5.2, 6) | 4.6 (4.3, 5.1) | 8.2 (7, 9.9) | 10.8 (9.9, 11.9) | 7.7 (7.2, 8.3) | 10.4 (9.4, 11.6) |
| White | ICU | 1.2 (1.1, 1.3) | 1.2 (1.1, 1.3) | 1.3 (1.1, 1.6) | 1.1 (0.9, 1.4) | 1.4 (1.1, 1.9) | 1.2 (1, 1.3) | 1 (0.9, 1.1) | 0.9 (0.8, 1) |
| **Number of comorbidities** | | | | | | | | | |
| 0 | Death | 7.2 (6.2, 8.1) | 7.1 (6.3, 7.9) | 8.7 (7.6, 10.1) | 10.7 (8.8, 12.5) | 9.5 (7.6, 11.9) | 9.8 (8.5, 11.1) | 10 (8.7, 11.5) | 12.4 (10.7, 14.5) |
| 0 | Discharge | 3.9 (3.7, 4.2) | 4 (3.8, 4.2) | 3.2 (3, 3.5) | 2.7 (2.4, 2.9) | 5.2 (4.4, 6.1) | 6.9 (6.2, 7.5) | 4.9 (4.5, 5.4) | 6.7 (6.1, 7.5) |
| 0 | ICU | 1.1 (1, 1.2) | 1.1 (1, 1.2) | 1.2 (1, 1.5) | 1 (0.8, 1.3) | 1.3 (1, 1.6) | 1.1 (0.9, 1.3) | 1 (0.8, 1.1) | 0.8 (0.7, 0.9) |
| 1 | Death | 7.2 (6.4, 7.9) | 7.1 (6.5, 7.7) | 8.7 (7.8, 9.8) | 10.7 (8.9, 12.4) | 9.5 (7.7, 11.8) | 9.7 (8.7, 10.9) | 10 (8.9, 11.1) | 12.4 (10.8, 14.1) |
| 1 | Discharge | 5 (4.6, 5.4) | 5 (4.7, 5.3) | 4 (3.7, 4.4) | 3.4 (3, 3.7) | 6.5 (5.5, 7.8) | 8.6 (7.9, 9.5) | 6.2 (5.6, 6.7) | 8.5 (7.7, 9.5) |
| 1 | ICU | 1.1 (1, 1.2) | 1.1 (1, 1.2) | 1.2 (1.1, 1.5) | 1 (0.8, 1.3) | 1.3 (1, 1.6) | 1.1 (1, 1.3) | 1 (0.8, 1.1) | 0.8 (0.7, 0.9) |
| 2 | Death | 6.9 (6.3, 7.6) | 6.9 (6.3, 7.4) | 8.5 (7.6, 9.5) | 10.4 (8.7, 12.2) | 9.2 (7.5, 11.5) | 9.5 (8.5, 10.6) | 9.7 (8.8, 10.7) | 12.1 (10.6, 13.8) |
| 2 | Discharge | 6.7 (6.1, 7.2) | 6.7 (6.3, 7.2) | 5.4 (5, 5.9) | 4.5 (4, 5) | 8.8 (7.3, 10.5) | 11.6 (10.5, 12.8) | 8.3 (7.6, 9.1) | 11.4 (10.2, 12.8) |
| 2 | ICU | 1.1 (1, 1.3) | 1.2 (1.1, 1.3) | 1.3 (1.1, 1.6) | 1.1 (0.9, 1.4) | 1.3 (1.1, 1.7) | 1.1 (1, 1.3) | 1 (0.9, 1.1) | 0.8 (0.7, 1) |
| 3+ | Death | 6.3 (5.9, 6.8) | 6.3 (5.9, 6.6) | 7.8 (7, 8.6) | 9.6 (8.1, 11.1) | 8.4 (6.8, 10.4) | 8.7 (7.9, 9.7) | 8.9 (8.2, 9.7) | 11.1 (9.9, 12.7) |
| 3+ | Discharge | 8.3 (7.8, 9) | 8.4 (8, 8.9) | 6.8 (6.3, 7.3) | 5.6 (5.1, 6.2) | 11 (9.3, 12.9) | 14.5 (13.2, 15.9) | 10.4 (9.6, 11.3) | 14.3 (12.8, 16) |
| 3+ | ICU | 1.2 (1.1, 1.3) | 1.2 (1.1, 1.4) | 1.4 (1.1, 1.6) | 1.2 (0.9, 1.4) | 1.4 (1.1, 1.8) | 1.2 (1.1, 1.4) | 1.1 (0.9, 1.2) | 0.9 (0.8, 1) |

Supplementary table 7: Estimated median time (in days) to next event from ICU admission and 95% confidence interval, by baseline characteristic.

| Group | Event | Mar | Apr | May | Jun/Jul/Aug | Sep | Oct | Nov | Dec |
|---|---|---|---|---|---|---|---|---|---|
| All | Death | 9.4 (8.4, 10.6) | 8.6 (7.8, 9.5) | 10.2 (8.1, 13.3) | 12.5 (8.6, 16.8) | 11 (7.2, 15.8) | 10.7 (9.1, 12.6) | 11.3 (9.4, 13.5) | 9.9 (7.5, 13.1) |
| All | Discharge | 24.4 (21.1, 28.5) | 23.8 (21.2, 26.6) | 18.3 (14.3, 23.8) | 13.8 (10.2, 18.5) | 18.9 (13.6, 26.3) | 16.8 (14.1, 20.3) | 22.8 (18.8, 26.6) | 38 (29, 48.9) |
| **Sex** | | | | | | | | | |
| Female | Death | 8.9 (7.4, 10.4) | 8.1 (6.9, 9.5) | 9.5 (7.2, 12.2) | 12 (8.5, 17.8) | 10.3 (6.8, 15.3) | 10.1 (8.1, 12.4) | 10.7 (8.7, 12.9) | 9.3 (6.6, 12.8) |
| Female | Discharge | 21.4 (18.2, 25.6) | 20.6 (17.6, 23.7) | 16.1 (12.2, 20.7) | 12.1 (8.9, 16.5) | 16.7 (11.2, 23.5) | 14.5 (11.6, 17.6) | 20.1 (16.4, 24.6) | 34.1 (26.6, 42.6) |
| Male | Death | 9.6 (8.5, 10.8) | 8.8 (7.9, 9.7) | 10.2 (7.8, 13.3) | 12.9 (9.1, 18.7) | 11.1 (7.7, 16.2) | 10.8 (9.1, 12.9) | 11.6 (9.7, 13.6) | 10.1 (7.5, 13.8) |
| Male | Discharge | 26.6 (22.6, 31.3) | 25.5 (22.6, 28.6) | 20.1 (15.5, 25.1) | 15.1 (11.3, 20.5) | 20.8 (13.9, 30.3) | 18 (14.5, 21.9) | 24.9 (20.7, 30.7) | 42.3 (32.9, 54.2) |
| **Age group** | | | | | | | | | |
| 15-44 | Death | 10.7 (8, 14.2) | 9.4 (7.1, 12.3) | 11.1 (7.9, 16.3) | 14.9 (9.6, 22.9) | 14 (8.7, 22.5) | 12.3 (8.8, 16.6) | 13.5 (10.1, 18.7) | 11.5 (7.8, 17.5) |
| 15-44 | Discharge | 16.4 (13.4, 20) | 15.8 (13.3, 19.2) | 11.8 (8.9, 15.7) | 9.3 (6.8, 12.9) | 12.2 (8.4, 17.2) | 10.9 (8.7, 13.8) | 14.5 (11.7, 17.9) | 28.1 (21.6, 36.9) |
| 45-64 | Death | 10.9 (9.6, 12.5) | 9.6 (8.5, 10.6) | 11.3 (8.9, 15.1) | 15.2 (10.6, 21.6) | 14.2 (9.6, 21) | 12.4 (10.3, 14.9) | 13.7 (11.4, 16.3) | 11.7 (8.9, 15.9) |
| 45-64 | Discharge | 26.7 (22.8, 31.4) | 25.7 (22.9, 29.6) | 19.1 (15.2, 24) | 15.1 (11.2, 20.2) | 19.9 (13.8, 27.8) | 17.7 (14.4, 21.5) | 23.6 (19.8, 28.2) | 45.7 (35, 59.1) |
| 65-74 | Death | 9.4 (8.2, 10.9) | 8.3 (7.3, 9.3) | 9.8 (7.4, 13.4) | 13.2 (9, 19.2) | 12.3 (8.4, 18.3) | 10.8 (8.8, 12.9) | 11.9 (10.1, 14.2) | 10.1 (7.6, 13.9) |
| 65-74 | Discharge | 27.3 (22.4, 33.8) | 26.3 (22.1, 31.2) | 19.5 (15.1, 25.6) | 15.4 (11.1, 21.7) | 20.3 (14.1, 29) | 18.1 (14.2, 22.7) | 24.2 (19.9, 29.7) | 46.7 (35, 62.6) |
| 75+ | Death | 7 (5.9, 8.3) | 6.1 (5, 7.2) | 7.3 (5.3, 9.9) | 10 (7, 14.8) | 9.3 (6.2, 13.9) | 8.1 (6.4, 9.8) | 9 (7.3, 10.8) | 7.5 (5.4, 10.6) |
| 75+ | Discharge | 28.8 (21.5, 38.5) | 27.7 (21.3, 36.3) | 20.6 (14.7, 28.6) | 16.2 (11.2, 24) | 21.4 (13.9, 31.4) | 19.1 (13.6, 26.1) | 25.4 (18.9, 34.3) | 49.2 (33.6, 68.7) |
| **Region of residence** | | | | | | | | | |
| London/South of England | Death | 10.1 (8.8, 11.5) | 9 (8.1, 10.1) | 10.8 (8.2, 14.2) | 13.7 (9.3, 19.6) | 12 (7.8, 17.5) | 11.4 (9.1, 14) | 12.6 (10.3, 15.4) | 11 (8, 14.9) |
| London/South of England | Discharge | 25 (21.6, 28.9) | 24.4 (20.9, 27.8) | 18.8 (14.5, 24.6) | 14.3 (10.2, 19.5) | 19.7 (13.6, 28.6) | 17.8 (13.9, 22.3) | 23.2 (18.8, 28.5) | 40.8 (31, 54.1) |
| Midlands and East of England | Death | 8.7 (7.4, 10.2) | 7.8 (6.8, 9.1) | 9.3 (6.9, 12.5) | 11.9 (7.9, 17.5) | 10.4 (6.8, 15.3) | 9.9 (8.2, 11.9) | 10.9 (9, 13.3) | 9.6 (6.9, 12.9) |
| Midlands and East of England | Discharge | 23.7 (19.7, 28.4) | 23.1 (19.2, 27.9) | 17.8 (13.5, 23.1) | 13.5 (9.8, 18.7) | 18.7 (12.4, 27.8) | 16.8 (13.4, 20.8) | 22 (17.7, 26.8) | 38.6 (29.6, 49.5) |
| North of England | Death | 9.3 (7.9, 11.1) | 8.4 (7.2, 9.9) | 10 (7.5, 13.4) | 12.7 (8.4, 18.4) | 11.1 (7.4, 16.5) | 10.6 (8.7, 12.6) | 11.7 (9.8, 14.1) | 10.3 (7.7, 14.2) |
| North of England | Discharge | 23.7 (19.5, 29.4) | 23.1 (19.7, 27.2) | 17.8 (14, 23) | 13.5 (9.7, 18.8) | 18.7 (12.7, 26.7) | 16.8 (13.9, 20.7) | 22 (18.1, 26.8) | 38.6 (29.8, 50.9) |
| **Ethnicity** | | | | | | | | | |
| Asian | Death | 10.5 (8.8, 12.5) | 9.4 (7.9, 10.9) | 11.4 (8.3, 14.9) | 13.9 (9.3, 20.2) | 12.1 (8.3, 18.1) | 12.1 (9.6, 15.2) | 12.6 (10.3, 15.4) | 10.7 (7.7, 14.7) |
| Asian | Discharge | 21.6 (16.9, 26.7) | 21.1 (17.4, 25.4) | 15.6 (11.3, 21.1) | 11.3 (8, 16.2) | 16.8 (11.6, 24.1) | 15.3 (12, 19.8) | 19.8 (15.2, 25.5) | 33.7 (24.8, 46) |
| Black | Death | 9.7 (7.9, 12) | 8.7 (7, 10.5) | 10.5 (7.7, 14.2) | 12.9 (8.6, 19.1) | 11.2 (7.4, 17) | 11.2 (8.6, 14.5) | 11.7 (8.9, 15.2) | 9.9 (6.6, 13.7) |
| Black | Discharge | 27.3 (20.5, 34.1) | 26.7 (21.1, 34.3) | 19.7 (14.1, 27.9) | 14.3 (9.7, 20.2) | 21.2 (14.2, 31.2) | 19.3 (14.3, 26.4) | 25 (18.4, 33.9) | 42.6 (29.3, 61.3) |
| Mixed/Other | Death | 11.2 (8.5, 14.6) | 10 (7.5, 13.2) | 12.1 (8.7, 16.7) | 14.7 (9.5, 23) | 12.9 (8.3, 19.4) | 12.9 (9.4, 17.4) | 13.4 (9.9, 18.1) | 11.4 (7.6, 16.7) |
| Mixed/Other | Discharge | 29.6 (22.2, 39.2) | 29 (22.7, 37.8) | 21.4 (15, 29.9) | 15.6 (10.7, 22.1) | 23.1 (14.7, 34.7) | 21 (15.2, 28.1) | 27.2 (20.1, 36.7) | 46.3 (32, 65.4) |
| White | Death | 9 (8, 10.3) | 8 (7.1, 9.1) | 9.8 (7.5, 12.7) | 12 (8.1, 17.6) | 10.4 (7.3, 15.1) | 10.5 (8.7, 12.4) | 10.9 (9, 12.9) | 9.2 (6.8, 12.6) |
| White | Discharge | 24 (20.3, 28) | 23.5 (20.4, 26.8) | 17.3 (13.6, 22.1) | 12.6 (9.2, 17.5) | 18.7 (13.2, 26.2) | 17 (13.8, 20.7) | 22 (18.1, 27.1) | 37.5 (28.1, 48.5) |
| **Number of comorbidities** | | | | | | | | | |
| 0 | Death | 11.1 (9.5, 13) | 10.1 (8.7, 11.5) | 11.9 (8.7, 15.8) | 14.9 (10, 20.8) | 13.1 (8.8, 18.8) | 12.3 (10, 14.8) | 13.3 (11, 15.8) | 11.3 (8.2, 15.1) |
| 0 | Discharge | 26.1 (21.8, 31.6) | 25.8 (21.9, 29.8) | 20.4 (15.9, 26.4) | 15.1 (10.6, 21.2) | 20.9 (14.2, 29.3) | 18.2 (14.7, 22.2) | 24.6 (19.8, 29.6) | 42.9 (33.2, 57.6) |
| 1 | Death | 10.3 (8.7, 12.4) | 9.3 (8, 10.8) | 11 (8.3, 14.5) | 13.7 (9.3, 19.7) | 12.1 (8, 17.4) | 11.3 (9.3, 13.9) | 12.3 (9.9, 14.9) | 10.4 (7.7, 14.2) |
| 1 | Discharge | 20.7 (17.4, 24.8) | 20.5 (17.4, 24.1) | 16.2 (12.6, 20.8) | 12 (8.6, 16.6) | 16.6 (11.2, 24.4) | 14.4 (11.6, 17.6) | 19.5 (15.9, 23.7) | 34 (26.1, 45.4) |
| 2 | Death | 8.6 (7.4, 10.2) | 7.8 (6.6, 9.2) | 9.3 (6.9, 12.2) | 11.7 (7.7, 16.8) | 10.3 (6.8, 14.9) | 9.6 (7.7, 11.6) | 10.4 (8.5, 12.5) | 8.8 (6.5, 12.2) |
| 2 | Discharge | 23.3 (18.8, 29.3) | 23.1 (19.4, 27.6) | 18.2 (13.7, 23.5) | 13.5 (9.9, 18.7) | 18.7 (13, 27.1) | 16.2 (13.1, 20.4) | 22 (17.3, 27.3) | 38.3 (28.9, 50.3) |
| 3+ | Death | 8.7 (7.5, 10.1) | 7.8 (6.9, 9) | 9.4 (7, 12.5) | 11.8 (7.9, 17) | 10.4 (7, 14.8) | 9.7 (8.1, 11.4) | 10.5 (8.6, 12.5) | 8.9 (6.5, 11.9) |
| 3+ | Discharge | 26.2 (22.2, 32) | 26 (22.3, 30.2) | 20.5 (15.6, 26.3) | 15.2 (11, 21.5) | 21.1 (14.8, 30.5) | 18.3 (14.7, 22.9) | 24.7 (20.2, 30.2) | 43.1 (32.5, 59.1) |

Supplementary table 8: Estimated odds ratios and 95% confidence intervals for outcomes from hospital admission, by baseline characteristic/month of admission

| Group | Transition | Odds ratio |
|---|---|---|
| **Month of admission (ref. Mar)** | | |
| Apr | Death | 1.3 (1.1, 1.5) |
| May | Death | 1.6 (1.3, 2) |
| Jun/Jul/Aug | Death | 0.8 (0.6, 1.1) |
| Sep | Death | 0.7 (0.5, 0.9) |
| Oct | Death | 0.8 (0.7, 1) |
| Nov | Death | 1 (0.8, 1.1) |
| Dec | Death | 1.3 (1.1, 1.6) |
| **Sex (ref. Female)** | | |
| Male | Death | 0.6 (0.6, 0.7) |
| **Age group (ref. 15-45)** | | |
| 45-64 | Death | 2.3 (1, 4.9) |
| 65-74 | Death | 8.2 (3.8, 18) |
| 75+ | Death | 71.1 (32.7, 154.4) |
| **Region of residence (ref. London/South of England)** | | |
| Midlands and East of England | Death | 1.3 (1, 1.7) |
| North of England | Death | 0.8 (0.6, 1.1) |
| **Ethnicity (ref. Asian)** | | |
| Black | Death | 1.2 (0.8, 1.8) |
| Mixed/Other | Death | 0.6 (0.3, 1.1) |
| White | Death | 2.1 (1.5, 3) |
| **Number of comorbidities (ref. 0 comorbidities)** | | |
| 1 comorbidity | Death | 1.4 (0.9, 2.1) |
| 2 comorbidities | Death | 2.3 (1.5, 3.5) |
| 3+ comorbidities | Death | 5.2 (3.6, 7.5) |
| **Month of admission (ref. Mar)** | | |
| Apr | Discharge | 1.4 (1.2, 1.6) |
| May | Discharge | 2.8 (2.3, 3.4) |
| Jun/Jul/Aug | Discharge | 3 (2.4, 3.9) |
| Sep | Discharge | 1.1 (0.8, 1.5) |
| Oct | Discharge | 1.2 (1, 1.5) |
| Nov | Discharge | 1.4 (1.2, 1.7) |
| Dec | Discharge | 1.6 (1.3, 1.9) |
| **Sex (ref. Female)** | | |
| Male | Discharge | 0.5 (0.5, 0.6) |
| **Age group (ref. 15-45)** | | |
| 45-64 | Discharge | 0.5 (0.3, 0.6) |
| 65-74 | Discharge | 0.5 (0.3, 0.7) |
| 75+ | Discharge | 1.5 (1, 2.2) |
| **Region of residence (ref. London/South of England)** | | |
| Midlands and East of England | Discharge | 0.9 (0.7, 1.1) |
| North of England | Discharge | 0.7 (0.6, 1) |
| **Ethnicity (ref. Asian)** | | |
| Black | Discharge | 1.2 (0.9, 1.8) |
| Mixed/Other | Discharge | 1 (0.6, 1.6) |
| White | Discharge | 1.5 (1.1, 1.9) |
| **Number of comorbidities (ref. 0 comorbidities)** | | |
| 1 comorbidity | Discharge | 0.7 (0.5, 1) |
| 2 comorbidities | Discharge | 0.8 (0.6, 1.1) |
| 3+ comorbidities | Discharge | 1 (0.7, 1.3) |

Supplementary table 9: Estimated odds ratios and 95% confidence intervals for outcomes from ICU admission, by baseline characteristic/month of admission

| **Group** | **Transition** | **Odds ratio** |
|---|---|---|
| **Month of admission (ref. Mar)** | | |
| Apr | Discharge | 1 (0.8, 1.3) |
| May | Discharge | 2 (1.3, 3) |
| Jun/Jul/Aug | Discharge | 2.8 (1.7, 4.9) |
| Sep | Discharge | 1.8 (1, 3.1) |
| Oct | Discharge | 1.3 (1, 1.8) |
| Nov | Discharge | 1.5 (1.1, 2.1) |
| Dec | Discharge | 2.5 (1.6, 3.8) |
| **Sex (ref. Female)** | | |
| Male | Discharge | 0.4 (0.3, 0.7) |
| **Age group (ref. 15-45)** | | |
| 45-64 | Discharge | 0.4 (0.2, 0.7) |
| 65-74 | Discharge | 0.2 (0.1, 0.3) |
| 75+ | Discharge | 0.1 (0, 0.2) |
| **Region of residence (ref. London/South of England)** | | |
| Midlands and East of England | Discharge | 0.7 (0.4, 1) |
| North of England | Discharge | 1 (0.6, 1.6) |
| **Ethnicity (ref. Asian)** | | |
| Black | Discharge | 1.2 (0.8, 1.7) |
| Mixed/Other | Discharge | 1.8 (1.2, 2.8) |
| White | Discharge | 1.2 (0.9, 1.5) |
| **Number of comorbidities (ref. 0 comorbidities)** | | |
| 1 comorbidity | Discharge | 1 (0.6, 1.6) |
| 2 comorbidities | Discharge | 0.5 (0.3, 0.9) |
| 3+ comorbidities | Discharge | 0.7 (0.4, 1.2) |

Supplementary table 10: Estimated odds ratios and 95% confidence intervals for outcomes from hospital admission, by baseline characteristic and month of admission

| Group | Transition | Apr | May | Jun/Jul/Aug | September | October | November | December |
|---|---|---|---|---|---|---|---|---|
| **Age group (ref. 15-45, Mar)** | | | | | | | | |
| 45-64 | Death | 0.9 (0.4, 2.5) | 1.3 (0.2, 7.7) | 0.7 (0.1, 5.4) | 0.9 (0.1, 12.9) | 1.1 (0.2, 6.4) | 1.6 (0.3, 8.3) | 0.9 (0.2, 4.2) |
| 65-74 | Death | 1.2 (0.4, 3.1) | 1.8 (0.3, 11.1) | 1.4 (0.2, 10) | 1.3 (0.1, 17.8) | 1.1 (0.2, 6.7) | 1.4 (0.3, 7) | 1.6 (0.3, 7.3) |
| 75+ | Death | 1.8 (0.7, 4.9) | 1.3 (0.2, 8) | 1.1 (0.2, 7.5) | 0.9 (0.1, 11.9) | 0.9 (0.2, 5.5) | 1.2 (0.2, 6.1) | 1.9 (0.4, 8.6) |
| **Region of residence (ref. London/South of England, Mar)** | | | | | | | | |
| Midlands and East of England | Death | 1.7 (1.2, 2.4) | 1.5 (0.9, 2.7) | 1.3 (0.6, 3.1) | 0.8 (0.2, 2.7) | 0.5 (0.2, 0.9) | 1.1 (0.6, 1.8) | 1.3 (0.8, 2.2) |
| North of England | Death | 1.3 (0.9, 1.9) | 1.6 (0.9, 2.8) | 2.2 (0.9, 5.4) | 1 (0.3, 3) | 1.1 (0.5, 2.1) | 1.4 (0.8, 2.4) | 1.1 (0.6, 1.9) |
| **Ethnicity (ref. Asian, Mar)** | | | | | | | | |
| Black | Death | 1 (0.5, 1.7) | 0.3 (0.1, 1.3) | 0.4 (0, 8.3) | 0.6 (0.1, 5.3) | 1.1 (0.2, 5.7) | 2.9 (0.7, 12.1) | 1.6 (0.4, 7.2) |
| Mixed/Other | Death | 1 (0.5, 2.2) | 2 (0.5, 8.5) | 0.6 (0, 38.2) | 1.3 (0.1, 14.2) | 0.6 (0.1, 2.9) | 0.6 (0.1, 3.1) | 0.6 (0.1, 3.4) |
| White | Death | 1.6 (1, 2.4) | 2.1 (0.9, 5.2) | 2.1 (0.7, 6.4) | 0.9 (0.3, 2.6) | 0.7 (0.4, 1.3) | 1.7 (0.8, 3.3) | 1.6 (0.7, 3.5) |
| **Number of comorbidities (ref. 0 comorbidities, Mar)** | | | | | | | | |
| 1 comorbidity | Death | 1.3 (0.8, 2.2) | 0.8 (0.4, 1.8) | 0.8 (0.2, 2.5) | 2.1 (0.6, 7.5) | 1.1 (0.5, 2.2) | 1.1 (0.6, 2.2) | 1.5 (0.8, 3) |
| 2 comorbidities | Death | 1.4 (0.9, 2.3) | 0.9 (0.4, 2) | 1.3 (0.4, 4.1) | 1 (0.3, 3.3) | 1 (0.5, 2) | 1.2 (0.6, 2.3) | 1.2 (0.6, 2.4) |
| 3+ comorbidities | Death | 1.1 (0.7, 1.8) | 1 (0.5, 2.1) | 0.6 (0.2, 1.5) | 0.6 (0.2, 1.7) | 0.9 (0.5, 1.8) | 1 (0.5, 1.7) | 1.1 (0.6, 2.1) |
| **Age group (ref. 15-45, Mar)** | | | | | | | | |
| 45-64 | Discharge | 1.1 (0.7, 1.7) | 0.6 (0.3, 1.2) | 1.2 (0.6, 2.5) | 0.7 (0.3, 1.6) | 0.8 (0.4, 1.4) | 0.7 (0.4, 1.3) | 1.4 (0.8, 2.3) |
| 65-74 | Discharge | 1.3 (0.8, 2) | 1 (0.4, 2.1) | 2 (0.9, 4.6) | 0.7 (0.3, 1.9) | 0.6 (0.3, 1.2) | 0.7 (0.4, 1.2) | 1.8 (1, 3.2) |
| 75+ | Discharge | 2.6 (1.5, 4.3) | 0.9 (0.4, 2) | 1.4 (0.6, 3.4) | 0.5 (0.2, 1.5) | 0.7 (0.4, 1.4) | 0.7 (0.4, 1.4) | 2.2 (1.1, 4.2) |
| **Region of residence (ref. London/South of England, Mar)** | | | | | | | | |
| Midlands and East of England | Discharge | 1.3 (1, 1.8) | 1.1 (0.6, 1.7) | 0.5 (0.3, 0.9) | 1.3 (0.5, 3.7) | 0.9 (0.5, 1.6) | 1.2 (0.8, 1.9) | 1.3 (0.8, 2) |
| North of England | Discharge | 0.9 (0.7, 1.3) | 1.2 (0.7, 2) | 0.7 (0.3, 1.4) | 0.9 (0.3, 2.2) | 1 (0.6, 1.9) | 0.8 (0.5, 1.3) | 0.6 (0.4, 1) |
| **Ethnicity (ref. Asian, Mar)** | | | | | | | | |
| Black | Discharge | 1.2 (0.8, 2) | 0.4 (0.2, 1) | 2 (0.5, 7.6) | 0.4 (0.1, 1.9) | 1.4 (0.4, 5.4) | 0.8 (0.2, 2.8) | 1.1 (0.3, 3.3) |
| Mixed/Other | Discharge | 1 (0.6, 1.8) | 0.8 (0.3, 2.3) | 2.3 (0.4, 12.2) | 1 (0.2, 5.1) | 0.6 (0.2, 1.8) | 0.7 (0.3, 1.7) | 0.8 (0.3, 2) |
| White | Discharge | 1.4 (1, 2) | 0.8 (0.4, 1.5) | 0.8 (0.4, 1.7) | 0.7 (0.3, 1.6) | 1 (0.6, 1.7) | 1.1 (0.6, 1.7) | 0.8 (0.4, 1.4) |
| **Number of comorbidities (ref. 0 comorbidities, Mar)** | | | | | | | | |
| 1 comorbidity | Discharge | 1.4 (1, 2.1) | 0.9 (0.5, 1.7) | 1 (0.4, 2.2) | 1.5 (0.6, 3.8) | 1.1 (0.7, 1.8) | 1 (0.6, 1.6) | 1.5 (0.9, 2.5) |
| 2 comorbidities | Discharge | 1.4 (1, 2.1) | 1 (0.6, 1.9) | 0.8 (0.4, 1.9) | 0.6 (0.2, 1.4) | 0.9 (0.5, 1.5) | 1 (0.6, 1.7) | 1.1 (0.6, 1.8) |
| 3+ comorbidities | Discharge | 1.7 (1.2, 2.4) | 1.4 (0.8, 2.5) | 0.6 (0.3, 1.2) | 0.5 (0.2, 1) | 0.9 (0.6, 1.4) | 1 (0.6, 1.5) | 1.5 (0.9, 2.4) |

Supplementary table 11: Estimated odds ratios and 95% confidence intervals for outcomes from ICU admission, by baseline characteristic and month of admission

| Group | Transition | Apr | May | Jun/Jul/Aug | September | October | November | December |
|---|---|---|---|---|---|---|---|---|
| **Sex (ref. Female, Mar)** | | | | | | | | |
| Male | Discharge | 1.6 (1, 2.8) | 1.5 (0.6, 3.7) | 1.2 (0.4, 4) | 1 (0.3, 3.5) | 0.8 (0.4, 1.7) | 1.7 (0.9, 3.2) | 1.9 (0.8, 4.7) |
| **Age group (ref. 15-45, Mar)** | | | | | | | | |
| 45-64 | Discharge | 0.5 (0.2, 1.4) | 1 (0.1, 7.8) | 1.6 (0.2, 13.5) | 3.2 (0.1, 70.3) | 1.4 (0.3, 6.6) | 0.9 (0.2, 3.8) | 0.9 (0.2, 4.3) |
| 65-74 | Discharge | 0.6 (0.2, 1.7) | 1.5 (0.2, 12.8) | 0.9 (0.1, 8.5) | 0.6 (0, 12.2) | 1 (0.2, 5.1) | 1.1 (0.2, 5.1) | 1.8 (0.3, 9.5) |
| 75+ | Discharge | 0.7 (0.2, 2.2) | 1.4 (0.1, 12.8) | 1.2 (0.1, 12.7) | 1.1 (0, 25.3) | 0.6 (0.1, 3.7) | 0.8 (0.2, 4.2) | 0.6 (0.1, 3.6) |
| **Region of residence (ref. London/South of England, Mar)** | | | | | | | | |
| Midlands and East of England | Discharge | 1 (0.6, 1.8) | 1 (0.4, 2.7) | 1.7 (0.4, 7.3) | 1.3 (0.1, 11.8) | 1.2 (0.3, 3.9) | 0.5 (0.2, 1.3) | 1 (0.4, 2.7) |
| North of England | Discharge | 1.7 (0.9, 3.1) | 1.8 (0.6, 5.3) | 1.1 (0.3, 5) | 1.1 (0.1, 8.6) | 1.6 (0.5, 5.3) | 0.6 (0.3, 1.6) | 1.2 (0.4, 3.7) |
| **Number of comorbidities (ref. 0 comorbidities, Mar)** | | | | | | | | |
| 1 comorbidity | Discharge | 0.8 (0.4, 1.5) | 0.7 (0.2, 2.5) | 1.5 (0.2, 11) | 0.9 (0.1, 5.5) | 1.3 (0.5, 3.5) | 1 (0.4, 2.4) | 0.9 (0.3, 2.8) |
| 2 comorbidities | Discharge | 1.3 (0.6, 2.6) | 1.4 (0.4, 5.4) | 3.5 (0.4, 29.6) | 1.2 (0.2, 6.7) | 1.9 (0.7, 4.8) | 1.3 (0.5, 3.3) | 2.6 (0.8, 9.1) |
| 3+ comorbidities | Discharge | 0.8 (0.4, 1.6) | 0.9 (0.3, 3.1) | 1 (0.2, 5.3) | 1.1 (0.2, 5.2) | 1.4 (0.6, 3.3) | 1.2 (0.5, 2.7) | 0.6 (0.2, 1.6) |

Supplementary table 12: Expected time ratios and 95% confidence intervals for outcomes from hospital admission, by baseline characteristic/month of admission

| Group | Transition | Expected time ratio |
|---|---|---|
| **Month of admission (ref. Mar)** | | |
| Apr | Death | 1 (0.9, 1.1) |
| May | Death | 1.2 (1.1, 1.3) |
| Jun/Jul/Aug | Death | 1.4 (1.2, 1.6) |
| Sep | Death | 1.3 (1.1, 1.5) |
| Oct | Death | 1.3 (1.1, 1.4) |
| Nov | Death | 1.3 (1.2, 1.4) |
| Dec | Death | 1.6 (1.4, 1.8) |
| **Sex (ref. Female)** | | |
| Male | Death | 1 (1, 1.1) |
| **Age group (ref. 15-45)** | | |
| 45-64 | Death | 1 (0.8, 1.3) |
| 65-74 | Death | 1 (0.7, 1.2) |
| 75+ | Death | 0.9 (0.7, 1.2) |
| **Region of residence (ref. London/South of England)** | | |
| Midlands and East of England | Death | 1 (0.9, 1.1) |
| North of England | Death | 1 (0.9, 1.1) |
| **Ethnicity (ref. Asian)** | | |
| Black | Death | 1 (0.9, 1.1) |
| Mixed/Other | Death | 1.1 (0.9, 1.4) |
| White | Death | 1.1 (1, 1.2) |
| **Number of comorbidities (ref. 0 comorbidities)** | | |
| 1 comorbidity | Death | 1 (0.9, 1.1) |
| 2 comorbidities | Death | 1 (0.9, 1.1) |
| 3+ comorbidities | Death | 0.9 (0.8, 1) |
| **Month of admission (ref. Mar)** | | |
| Apr | Discharge | 1 (0.9, 1.1) |
| May | Discharge | 0.8 (0.7, 0.9) |
| Jun/Jul/Aug | Discharge | 0.7 (0.6, 0.7) |
| Sep | Discharge | 1.2 (1, 1.4) |
| Oct | Discharge | 1.7 (1.5, 1.9) |
| Nov | Discharge | 1.2 (1.1, 1.3) |
| Dec | Discharge | 1.5 (1.4, 1.7) |
| **Sex (ref. Female)** | | |
| Male | Discharge | 1.1 (1, 1.2) |
| **Age group (ref. 15-45)** | | |
| 45-64 | Discharge | 1.7 (1.6, 1.8) |
| 65-74 | Discharge | 2.4 (2.2, 2.6) |
| 75+ | Discharge | 3.7 (3.5, 4) |
| **Region of residence (ref. London/South of England)** | | |
| Midlands and East of England | Discharge | 1.3 (1.2, 1.3) |
| North of England | Discharge | 1.7 (1.6, 1.9) |
| **Ethnicity (ref. Asian)** | | |
| Black | Discharge | 1.1 (1, 1.3) |
| Mixed/Other | Discharge | 1.1 (0.9, 1.3) |
| White | Discharge | 1.6 (1.5, 1.7) |
| **Number of comorbidities (ref. 0 comorbidities)** | | |
| 1 comorbidity | Discharge | 1.3 (1.2, 1.4) |
| 2 comorbidities | Discharge | 1.7 (1.6, 1.8) |
| 3+ comorbidities | Discharge | 2.1 (2, 2.3) |
| **Month of admission (ref. Mar)** | | |
| Apr | ICU | 1 (0.9, 1.2) |
| May | ICU | 1.2 (1, 1.4) |
| Jun/Jul/Aug | ICU | 1 (0.8, 1.2) |
| Sep | ICU | 1.2 (0.9, 1.5) |
| Oct | ICU | 1 (0.9, 1.2) |
| Nov | ICU | 0.9 (0.8, 1) |
| Dec | ICU | 0.7 (0.6, 0.9) |
| **Sex (ref. Female)** | | |
| Male | ICU | 1 (0.9, 1.1) |
| **Age group (ref. 15-45)** | | |
| 45-64 | ICU | 1.2 (1, 1.4) |
| 65-74 | ICU | 1.2 (1, 1.4) |
| 75+ | ICU | 1.3 (1.1, 1.6) |
| **Region of residence (ref. London/South of England)** | | |
| Midlands and East of England | ICU | 1.2 (1.1, 1.4) |
| North of England | ICU | 1.1 (0.9, 1.2) |
| **Ethnicity (ref. Asian)** | | |
| Black | ICU | 1 (0.8, 1.2) |
| Mixed/Other | ICU | 0.8 (0.7, 1) |
| White | ICU | 1.1 (1, 1.2) |
| **Number of comorbidities (ref. 0 comorbidities)** | | |
| 1 comorbidity | ICU | 1 (0.9, 1.1) |
| 2 comorbidities | ICU | 1 (0.9, 1.2) |
| 3+ comorbidities | ICU | 1.1 (1, 1.2) |

Supplementary table 13: Expected time ratios and 95% confidence intervals for outcomes from ICU admission, by baseline characteristic/month of admission

| Group | Transition | Expected time ratio |
|---|---|---|
| **Month of admission (ref. Mar)** | | |
| Apr | Death | 0.9 (0.8, 1.1) |
| May | Death | 1.1 (0.8, 1.3) |
| Jun/Jul/Aug | Death | 1.3 (0.9, 1.7) |
| Sep | Death | 1.1 (0.8, 1.5) |
| Oct | Death | 1.1 (0.9, 1.3) |
| Nov | Death | 1.2 (1, 1.4) |
| Dec | Death | 1 (0.8, 1.3) |
| **Sex (ref. Female)** | | |
| Male | Death | 1.1 (1, 1.2) |
| **Age group (ref. 15-45)** | | |
| 45-64 | Death | 1 (0.8, 1.3) |
| 65-74 | Death | 0.9 (0.7, 1.1) |
| 75+ | Death | 0.7 (0.6, 0.9) |
| **Region of residence (ref. London/South of England)** | | |
| Midlands and East of England | Death | 0.9 (0.8, 1) |
| North of England | Death | 0.9 (0.8, 1.1) |
| Black | Death | 0.9 (0.8, 1.1) |
| Mixed/Other | Death | 1.1 (0.8, 1.3) |
| White | Death | 0.9 (0.8, 1) |
| **Number of comorbidities (ref. 0 comorbidities)** | | |
| 1 comorbidity | Death | 0.9 (0.8, 1.1) |
| 2 comorbidities | Death | 0.8 (0.7, 0.9) |
| 3+ comorbidities | Death | 0.8 (0.7, 0.9) |
| **Month of admission (ref. Mar)** | | |
| Apr | Discharge | 1 (0.8, 1.2) |
| May | Discharge | 0.8 (0.6, 1) |
| Jun/Jul/Aug | Discharge | 0.6 (0.4, 0.8) |
| Sep | Discharge | 0.8 (0.5, 1.1) |
| Oct | Discharge | 0.7 (0.5, 0.9) |
| Nov | Discharge | 0.9 (0.7, 1.2) |
| Dec | Discharge | 1.6 (1.2, 2.1) |
| **Sex (ref. Female)** | | |
| Male | Discharge | 1.2 (1.1, 1.4) |
| **Age group (ref. 15-45)** | | |
| 45-64 | Discharge | 1.6 (1.4, 1.9) |
| 65-74 | Discharge | 1.7 (1.3, 2.1) |
| 75+ | Discharge | 1.8 (1.3, 2.4) |
| **Region of residence (ref. London/South of England)** | | |
| Midlands and East of England | Discharge | 0.9 (0.8, 1.1) |
| North of England | Discharge | 0.9 (0.8, 1.1) |
| **Ethnicity (ref. Asian)** | | |
| Black | Discharge | 1.3 (0.9, 1.7) |
| Mixed/Other | Discharge | 1.4 (1, 1.9) |
| White | Discharge | 1.1 (0.9, 1.4) |
| **Number of comorbidities (ref. 0 comorbidities)** | | |
| 1 comorbidity | Discharge | 0.8 (0.7, 1) |
| 2 comorbidities | Discharge | 0.9 (0.7, 1.1) |
| 3+ comorbidities | Discharge | 1 (0.8, 1.2) |

Supplementary Figure 1: Goodness of fit for "From Hospital' sub-model, by month of admission (panel A), and sex (panel B), age group (panel C), region of residence (panel D), ethnicity (panel E), and number of comorbidities (panel F). Solid lines are Aalen-Johansen cumulative incidence curves, dotted lines are derived from the parametric mixture model.

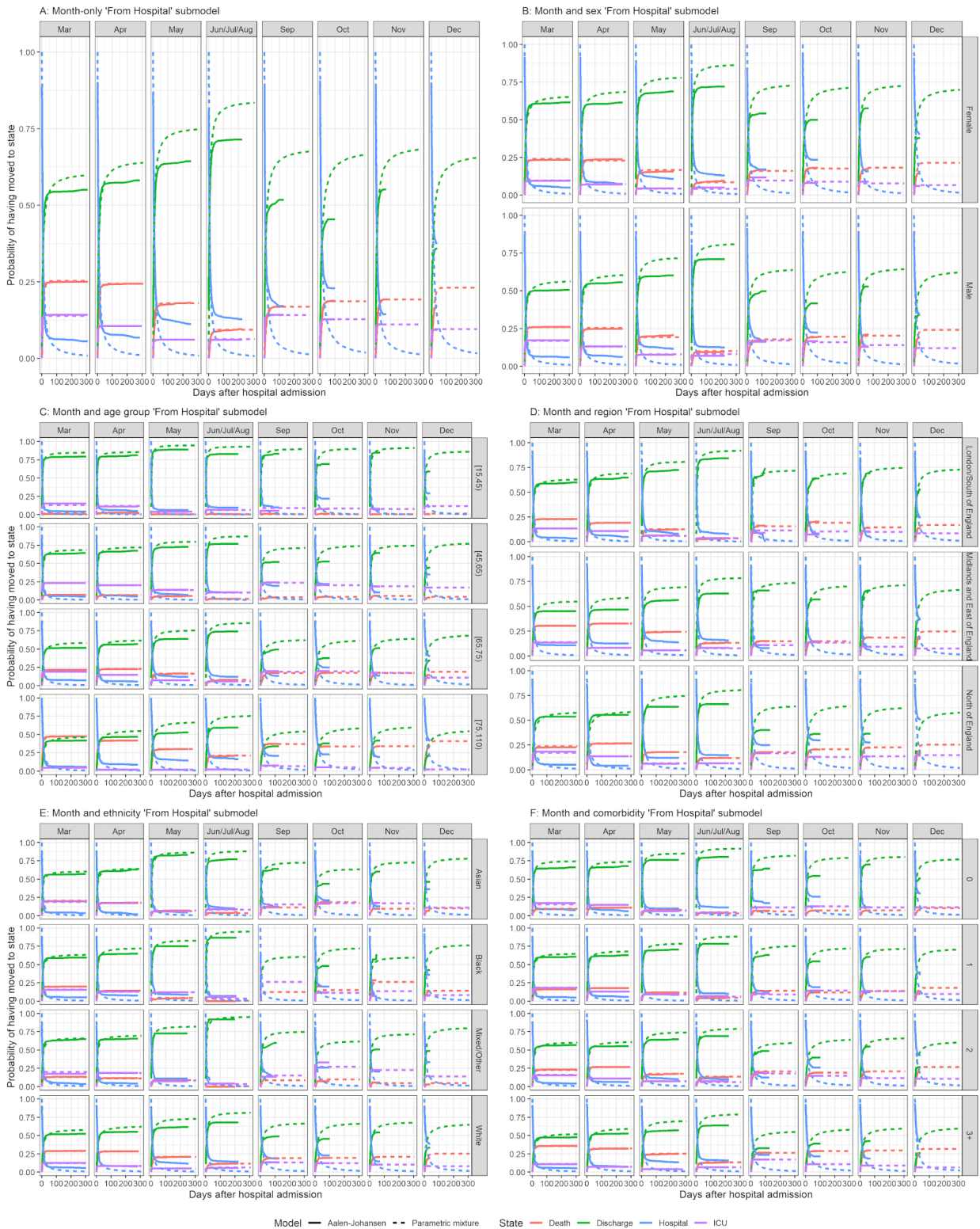

Supplementary Figure 2: Goodness of fit for "From ICU" sub-model, by month of admission (panel A), and sex (panel B), age group (panel C), region of residence (panel D), ethnicity (panel E), and number of comorbidities (panel F). Solid lines are Aalen-Johansen cumulative incidence curves, dotted lines are derived from the parametric mixture model.

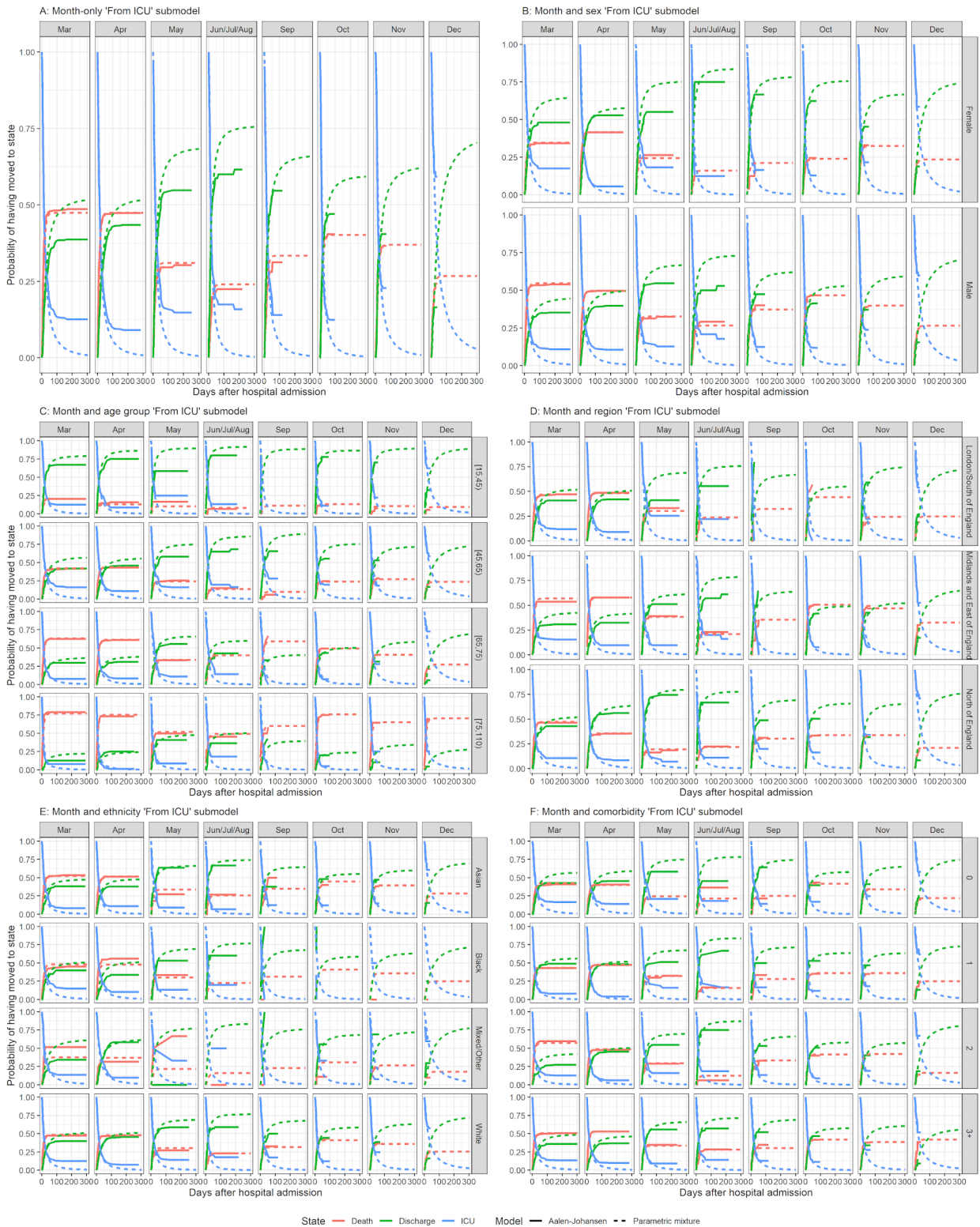

Supplementary Figure 3: Estimated hospitalised case-fatality risk (HFR) averaged over ICU and non-ICU admission. Error bars are 95% confidence intervals to represent uncertainty in the estimated probability.

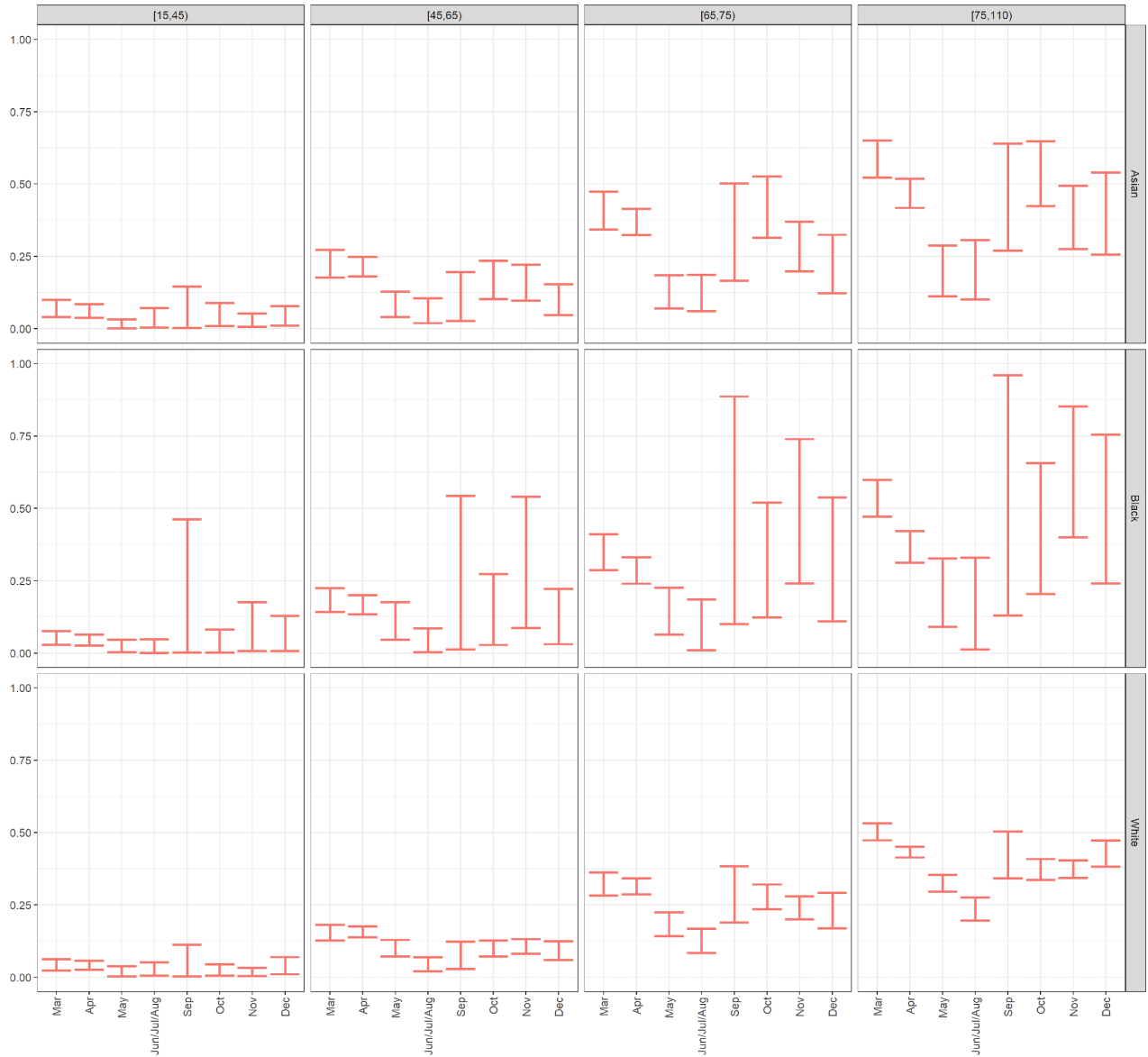

Supplementary Figure 4: Estimated hospitalised case-fatality risk (HFR) averaged over ICU and non-ICU admission, by month of admission, region of residence (line range) and regional adult critical care occupancy rate[1] (bar).

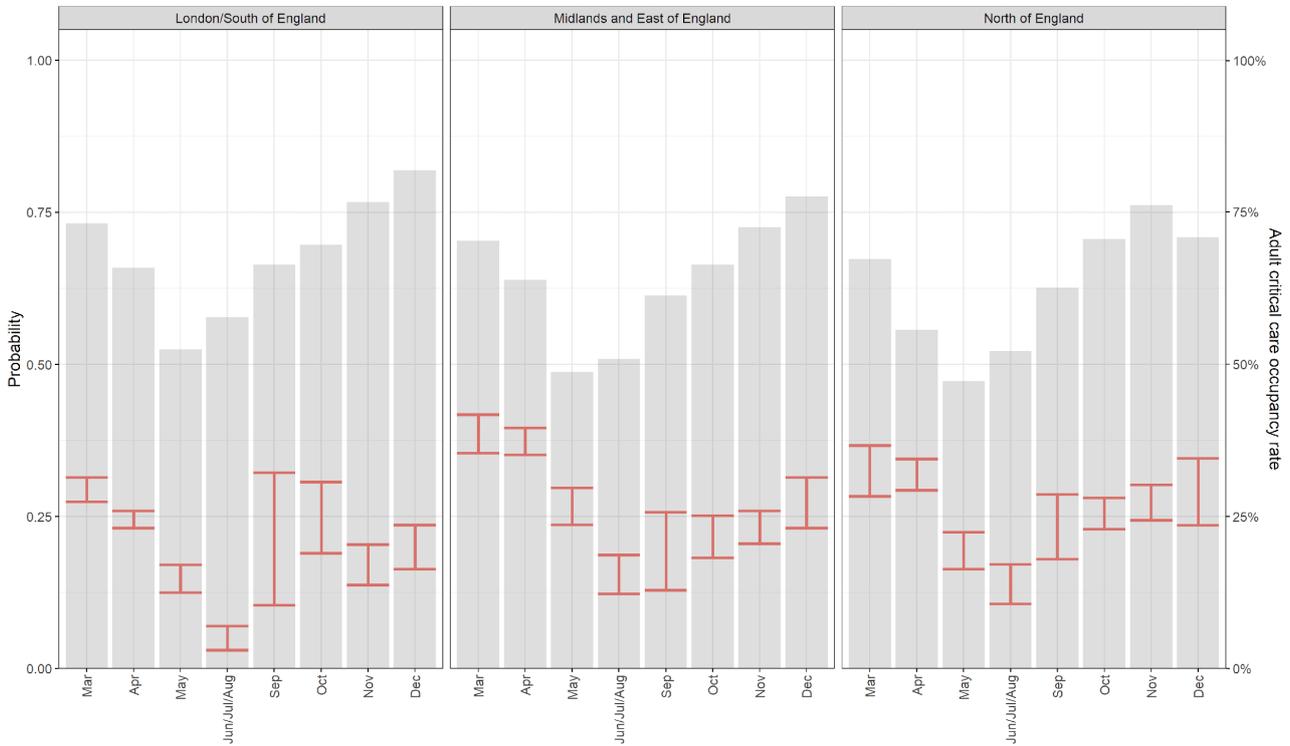

[1]Data on regional adult critical care occupancy rate accessed from:

https://www.england.nhs.uk/statistics/statistical-work-areas/uec-sitrep/urgent-and-emergency-care-daily-situation-reports-2020-21/